\documentclass[apj,iop,twocolappendix,numberedappendix]{emulateapj}
\usepackage{amsmath}
\usepackage{multirow}
\usepackage{relsize}

\newcommand{\ak}{A_\mathrm{K}}
\newcommand{\mstar}{M_\star}
\newcommand{\mgas}{M_\mathrm{gas}}
\newcommand{\Df}{D_\mathrm{f}}
\newcommand{\alphavircirc}{\alpha_\mathrm{vir,\circ}}
\newcommand{\alphavir}{\alpha_\mathrm{vir}}
\newcommand{\sfe}{\mathrm{SFE}}

\newcommand{\sigs}{\sigma_s}

\newcommand{\eps}{\epsilon}

\newcommand{\meanrho}{\rho_0}
\newcommand{\means}{s_0}
\newcommand{\deriv}{\,\mathrm{d}}
\newcommand{\mach}{\mathcal{M}}
\newcommand{\macha}{\mathcal{M}_\mathrm{A}}
\newcommand{\cs}{c_\mathrm{s}}
\newcommand{\va}{v_\mathrm{A}}

\newcommand{\mc}{M_\mathrm{c}}

\newcommand{\tsf}{t_\mathrm{SF}}

\newcommand{\ptwod}{P_\mathrm{2D}}
\newcommand{\pthreed}{P_\mathrm{3D}}
\newcommand{\dvar}{\sigma_\Delta^2}

\newcommand{\rhosink}{\rho_\mathrm{sink}}
\newcommand{\rsink}{r_\mathrm{sink}}

\newcommand{\msol}{\mbox{$M_{\sun}$}}
\newcommand{\g}{\mathrm{g}}
\newcommand{\cm}{\mathrm{cm}}
\newcommand{\km}{\mathrm{km}}
\newcommand{\pc}{\mathrm{pc}}

\newcommand{\s}{\mathrm{s}}
\newcommand{\yr}{\mathrm{yr}}
\newcommand{\Gauss}{\mathrm{G}}

\newcommand{\vect}[1]{{\mathbf{#1}}}

\slugcomment{The Astrophysical Journal, in press, \today}

\shorttitle{The Star Formation Efficiency}
\shortauthors{Federrath \& Klessen}

\begin{document}

\title{On the Star Formation Efficiency of Turbulent Magnetized Clouds}

\author{Christoph~Federrath\altaffilmark{1}, \& Ralf~S.~Klessen\altaffilmark{2}}
\email{christoph.federrath@monash.edu}

\altaffiltext{1}{Monash Centre for Astrophysics, School of Mathematical Sciences, Monash University, Vic 3800, Australia}
\altaffiltext{2}{Institut f\"ur Theoretische Astrophysik, Zentrum f\"ur Astronomie, Universit\"at Heidelberg, Albert-Ueberle-Str.~2, D-69120 Heidelberg, Germany}

\begin{abstract}
We study the star formation efficiency (SFE) in simulations and observations of turbulent, magnetized, molecular clouds. We find that the probability density functions (PDFs) of the density and the column density in our simulations with solenoidal, mixed, and compressive forcing of the turbulence, sonic Mach numbers of \mbox{$3$--$50$}, and magnetic fields in the super- to the trans-Alfv\'enic regime, all develop power-law tails of flattening slope with increasing SFE. The high-density tails of the PDFs are consistent with equivalent radial density profiles, $\rho\propto r^{-\kappa}$ with \mbox{$\kappa\sim1.5$--$2.5$}, in agreement with observations. Studying velocity--size scalings, we find that all the simulations are consistent with the observed \mbox{$v\propto\ell^{1/2}$} scaling of supersonic turbulence, and seem to approach Kolmogorov turbulence with $v\propto\ell^{1/3}$ below the sonic scale. The velocity--size scaling is, however, largely independent of the SFE. In contrast, the density--size and column density--size scalings are highly sensitive to star formation. We find that the power-law slope $\alpha$ of the density power spectrum, $\pthreed(\rho,k)\propto k^\alpha$, or equivalently the $\Delta$-variance spectrum of the column density, $\dvar(\Sigma,\ell)\propto\ell^{-\alpha}$, switches sign from $\alpha\lesssim0$ for $\sfe\sim0$ to $\alpha\gtrsim0$ when star formation proceeds ($\sfe>0$). We provide a relation to compute the SFE from a measurement of $\alpha$. Studying the literature, we find values ranging from $\alpha=-1.6$ to $+1.6$ in observations covering scales from the large-scale atomic medium, over cold molecular clouds, down to dense star-forming cores. From those $\alpha$ values, we infer SFEs and find good agreement with independent measurements based on young stellar object (YSO) counts, where available. Our \mbox{$\sfe$--$\alpha$} relation provides an independent estimate of the $\sfe$ based on the column density map of a cloud alone, without requiring a priori knowledge of star-formation activity or YSO counts.
\end{abstract}


\section{Introduction}

The most important physical processes determining star formation are turbulence, gravity, and magnetic fields \citep[see the reviews by][]{MacLowKlessen2004,ElmegreenScalo2004,ScaloElmegreen2004,McKeeOstriker2007}. On one hand, molecular cloud turbulence---because it is supersonic---compresses interstellar gas in shocks and filaments. If a critical amount of mass is swept up in a local compression, that gas can become gravitationally unstable and collapse, giving birth to new stars. On the other hand, turbulence and magnetic fields together carry about the same amount of energy as there is in gravitational binding energy in a whole molecular cloud \citep[e.g.,][]{StahlerPalla2004}. Thus, both turbulence and magnetic fields provide large-scale support against global cloud collapse. That support is needed, because otherwise molecular clouds would simply collapse as a whole, which is not observed \citep{ZuckermanPalmer1974,ZuckermanEvans1974}. The role of gravity and magnetic pressure is unidirectional: gravity always acts to promote collapse, while magnetic pressure always counteracts collapse. The role of turbulence on the other hand is dual: it provides global support, but at the same time produces local compression---the seeds for stellar birth.

Hence, turbulence, gravity, and magnetic fields are crucial for our understanding of star formation. These three physical effects are likely the major players in controlling the star formation rate (SFR) in the Milky Way and potentially in other galaxies, as discussed in \citet[][hereafter Paper I]{FederrathKlessen2012}. There, we compared theoretical and numerical models of the SFR with observations of Galactic clouds, finding very good agreement for a wide range of physical parameters, including extreme cases of turbulent forcing (solenoidal versus compressive forcing of the turbulence), sonic Mach numbers between 3 and 50, and magnetic fields ranging from the super-Alfv\'enic to the trans-Alfv\'enic regime, all in the range of parameters observed in real molecular clouds.

Despite this large variety of physical parameters, the conversion of gas into stars is always quite inefficient in molecular clouds, with SFEs of typically only a few percent \citep{MyersEtAl1986,EvansEtAl2009,LadaLombardiAlves2010}. However, it is still poorly understood why this is the case. It is the aim of this study to shed light on this subject by providing a detailed analysis of the SFE in the simulations of Paper I, and comparing them to observations of Galactic clouds.

The SFE of clouds is often uncertain or unknown. The main difficulty in estimating the SFE lies in obtaining a clear census of YSOs, in particular in dense and confused regions where many protostellar disks are embedded and closely packed, such that the YSO population is often underestimated in observations. There is some indication that the SFE ranges from less than 1\% up to 50\% in most clouds, depending on the particular cloud studied, its physical parameters, the actual subregion considered inside that cloud, and the evolutionary stage of that region. This is a very wide range of SFEs. The aim of this paper is to advance our understanding of the physical processes that determine this range and provide measures to constrain the SFE for a particular cloud region observed. For instance, considered as a whole, giant molecular clouds typically only convert at most a few percent of their gas into stars \citep[e.g.,][]{MyersEtAl1986,EvansEtAl2009}. In contrast, when looking at regions of dense core and cluster formation, observers find higher SFEs \citep[e.g.,][]{WilkingLada1983,BeutherEtAl2002,KoenigEtAl2008}, eventually approaching the local efficiency $\eps$ on scales of individual protostellar cores. This local efficiency $\eps$ determines the gas fraction of a single protostellar accretion envelope that actually ends up on the protostar(s) forming in a dense core. Although gas is efficiently falling onto the protostar, jets, winds and outflows launched from the protostellar disk counteract the inflow, thereby limiting accretion such that $\eps<1$. We estimated this local efficiency with \mbox{$\eps\sim30\%$--$70\%$} by comparing our simulations with Galactic observations by \citet{HeidermanEtAl2010} in Paper I. It is thus plausible that the SFE naturally approaches the local efficiency $\eps$ on sufficiently small scales, when a single, dense, star-forming core is considered.

Here, we investigate the dependence of the SFE on molecular cloud scales and on physical parameters, such as the driving of turbulence, the sonic and Alfv\'en Mach numbers, and the virial parameter of a cloud. First, we study the structure of the column density and how it changes with $\sfe$ in Section~\ref{sec:morphology}. We then discuss the dependences of the probability density functions (PDFs) of the volumetric and column densities in Section~\ref{sec:pdfs}. We find that the PDFs depend on the forcing of the turbulence and on the sonic and Alfv\'en Mach numbers. The PDFs develop power-law tails of flatting slope in all our numerical models when star formation proceeds, as seen in observations of clouds \citep{KainulainenEtAl2009,SchneiderEtAl2012}, suggesting that the PDF can be used to infer the SFE. In Section~\ref{sec:scaling}, however, we find that a more reliable and independent distinction between star-forming and quiescent clouds is possible by studying the spatial scaling of column density with Fourier or $\Delta$-variance analyses \citep{OssenkopfKlessenHeitsch2001}, i.e., by studying the density--size scaling in molecular clouds. We extend this qualitative discriminator of star formation to a quantitative method for estimating the SFE in Section~\ref{sec:estimatingsfe}, by measuring how the slope $\alpha$ of the density power spectrum changes with $\sfe$, depending on the forcing of turbulence, the sonic Mach number, and the magnetic field. Inverting this relation yields a function $\sfe(\alpha)$, which can be used to estimate $\sfe$ from a measurement of $\alpha$ in a dust or integrated molecular line map. The advantage of this new method is that no a priori knowledge of star formation activity or YSO counts is required to estimate SFE. In Section~\ref{sec:obs} we apply this method to observations and compare inferred SFEs with independent estimates, indicating good agreement. In Section~\ref{sec:uncertainties}, we discuss uncertainties and limitations of the new method. Our conclusions are summarized in Section~\ref{sec:conclusions}.

\section{Numerical Simulations} \label{sec:sims}

Our numerical simulation techniques are explained in detail in Paper I. Here we only give a brief overview of the most important aspects of the simulation methodology and provide a complete list of simulation parameters as in Paper I.

We use the adaptive mesh refinement \citep[AMR,][]{BergerColella1989} code FLASH v2.5 \citep{FryxellEtAl2000,DubeyEtAl2008} to model isothermal, self-gravitating, magnetohydrodynamic (MHD) turbulence on three-dimensional (3D), periodic grids with resolutions of $N_\mathrm{res}^3=128^3$--$1024^3$ grid points. These are all uniform-grid simulations, except for one $N_\mathrm{res}^3=1024^3$ simulation, where we use a root grid with $512^3$ cells and one level of AMR with a refinement criterion to ensure that the local Jeans lengths is covered with at least 32 grid cells, in order to resolve turbulent vorticity and magnetic-field amplification on the Jeans scale \citep{SurEtAl2010,FederrathSurSchleicherBanerjeeKlessen2011,TurkEtAl2012}. For solving the MHD equations, we use the HLL3R positive-definite Riemann solver \citep{WaaganFederrathKlingenberg2011}. The MHD equations are closed with an isothermal equation of state, which is a reasonable approximation for dense, molecular gas of solar metallicity, over a wide range of densities \citep{WolfireEtAl1995,OmukaiEtAl2005,PavlovskiSmithMacLow2006,GloverMacLow2007a,GloverMacLow2007b,GloverFederrathMacLowKlessen2010,HillEtAl2011,HennemannEtAl2012}. The self-gravity of the gas is computed with a multi-grid Poisson solver \citep[see][]{Ricker2008}.

\subsection{Turbulent Forcing} \label{sec:forcing}

To drive turbulence in the simulations, we apply a stochastic acceleration field ${\bf F_\mathrm{stir}}$ as a momentum and energy source term. ${\bf F_\mathrm{stir}}$ only contains large-scale modes, $1<k<3$, where most of the power is injected at the $k=2$ mode in Fourier space, which corresponds to half of the box size $L$ in physical space. We thus model turbulent forcing on large scales, as favored by molecular cloud observations \citep[e.g.,][]{OssenkopfMacLow2002,HeyerWilliamsBrunt2006,BruntHeyerMacLow2009,RomanDuvalEtAl2011}. Smaller scales, $k\geq3$, are not affected directly by the forcing, such that turbulence can develop self-consistently there. We use the Ornstein-Uhlenbeck (OU) process to model ${\bf F_\mathrm{stir}}$, which is a well-defined stochastic process with a finite autocorrelation timescale \citep{EswaranPope1988,SchmidtHillebrandtNiemeyer2006}. We set the autocorrelation time equal to the turbulent crossing time on the largest scales of the system, $T=L/(2\mach\cs)$, with the RMS sonic Mach number $\mach$ and the constant sound speed $\cs=0.2\,\km\,\s^{-1}$, leading to a smoothly varying stochastic forcing in space and time \citep[for details, see][Paper I]{SchmidtEtAl2009,FederrathDuvalKlessenSchmidtMacLow2010,KonstandinEtAl2012}.

We can adjust the mixture of solenoidal and compressive modes of our turbulent forcing, by applying a projection in Fourier space. Here we compare simulations with solenoidal forcing ($\nabla\cdot{\bf F_\mathrm{stir}}=0$) and compressive forcing ($\nabla\times{\bf F_\mathrm{stir}}=0$), as well as an intermediate mixture of both. Solenoidal and compressive forcing are extreme cases, while in real molecular clouds we expect some mixture with a range of possible ratios between solenoidal and compressive modes (see the discussion of physical drivers of turbulence and their mode characteristics in Paper I). For instance, \citet{MotteAndreNeri1998} describe a scenario in which the observed alignment of young protostars in the $\rho$ Ophiuchi central region could have been triggered by a shock compression, induced by a supernova/wind shell and/or expanding $\mathrm{H}\,\textsc{ii}$ regions from nearby OB associations. Such expanding shells are compressive forcing mechanisms, because they primarily excite compressible velocity modes.

\subsection{Sink Particles} \label{sec:sinks}
In order to model collapse and accretion, we use an advanced AMR-based approach for sink particles, in which only bound and collapsing gas is accreted \citep[for a detailed analysis, see][]{FederrathBanerjeeClarkKlessen2010}. The key feature of this approach is to define a control volume centered on grid cells exceeding a density threshold. \citet{TrueloveEtAl1997} found that the Jeans length must be resolved with at least 4 grid cells to avoid artificial fragmentation, leading to a resolution-dependent density threshold criterion for triggering sink particle creation checks,
\begin{equation} \label{eq:rhosink}
\rhosink = \frac{\pi\cs^2}{4G\,\rsink^2}\,,
\end{equation}
where the sink particle accretion radius $\rsink$ is set to 2.5 grid-cell lengths at the maximum level of refinement, corresponding to half a Jeans length at $\rhosink$. This guarantees that the Jeans length is still resolved with 5 grid cells prior to potential sink particle creation to avoid artificial fragmentation.

Grid cells exceeding the density threshold given by Equation~(\ref{eq:rhosink}), however, do not form sink particles right away. First, a spherical control volume with radius $\rsink$ is defined around the cell exceeding $\rhosink$, in which a series of checks for gravitational instability and collapse are performed. If all checks are passed, a sink particle is created in the center of the control volume. This procedure avoids spurious sink particle formation, and allows us to trace only truly collapsing and star-forming gas.

Once a sink particle is created, it can gain mass by accreting gas from the AMR grid, but only if this gas exceeds the threshold density, is inside the sink particle accretion radius, is bound to the particle, and is collapsing toward it. If all these criteria are fulfilled, the excess mass above the density threshold defined by Equation~(\ref{eq:rhosink}) is removed from the MHD system and added to the sink particle, such that mass, momentum and angular momentum are conserved by construction \citep[see][for details]{FederrathBanerjeeClarkKlessen2010,FederrathBanerjeeSeifriedClarkKlessen2011}.

Sink particle gravity is computed by direct $N$-body summation over all sink particles and grid cells. We use a second-order accurate Leapfrog integrator to advance the sink particles on a timestep that allows us to resolve close and highly eccentric orbits of sink particles without introducing significant errors on super-resolution grid scales, as tested in \citet{FederrathBanerjeeClarkKlessen2010}.

\begin{table*}
\caption{Basic Parameters of the Numerical Models of Forced, Supersonic, Self-gravitating, (Magneto)hydrodynamic Turbulence.}
\label{tab:sims}
\def\arraystretch{1.2}
\setlength{\tabcolsep}{3.0pt}
\begin{tabular}{l|rcccccccc|ccccc}
\hline
\hline
Model & $N_\mathrm{res}$ & Forcing & $\meanrho[\g\,\cm^{-3}]$ & $L[\pc]$ & $M_c[\msol]$ & $\sigma_V[\km/\s]$ & $B_0[\mu\Gauss]$ & $\beta_0$ & $\alphavircirc$ & $\alphavir$ & $\mach$ & $b$ & $\beta$ & $\macha$\\
(1) & (2) & (3) & (4) & (5) & (6) & (7) & (8) & (9) & (10) & (11) & (12) & (13) & (14) & (15) \\
\hline
01) GT256sM3                       & 256  & sol  & $  5.8\!\times\!10^{-19}$ & $  3.3\!\times\!10^{-1}$ & $  3.1\!\times\!10^{2}$ & $0.59$ & $ 0$ & $\infty$ & $0.07$ & $ 1.4$ & $ 2.9$ & $1/3$ & $\infty$ & $\infty$ \\
02) GT512sM3                       & 512  & sol  & $  5.8\!\times\!10^{-19}$ & $  3.3\!\times\!10^{-1}$ & $  3.1\!\times\!10^{2}$ & $0.59$ & $ 0$ & $\infty$ & $0.07$ & $ 1.4$ & $ 3.0$ & $1/3$ & $\infty$ & $\infty$ \\
03) GT256mM3                       & 256  & mix  & $  5.8\!\times\!10^{-19}$ & $  3.3\!\times\!10^{-1}$ & $  3.1\!\times\!10^{2}$ & $0.61$ & $ 0$ & $\infty$ & $0.08$ & $ 1.1$ & $ 3.1$ & $0.4$ & $\infty$ & $\infty$ \\
04) GT256cM3                       & 256  & comp & $  5.8\!\times\!10^{-19}$ & $  3.3\!\times\!10^{-1}$ & $  3.1\!\times\!10^{2}$ & $0.58$ & $ 0$ & $\infty$ & $0.07$ & $0.46$ & $ 2.9$ & $1$   & $\infty$ & $\infty$ \\
05) GT512cM3                       & 512  & comp & $  5.8\!\times\!10^{-19}$ & $  3.3\!\times\!10^{-1}$ & $  3.1\!\times\!10^{2}$ & $0.58$ & $ 0$ & $\infty$ & $0.07$ & $0.48$ & $ 2.9$ & $1$   & $\infty$ & $\infty$ \\
\hline
06) GT256sM5                       & 256  & sol  & $  3.3\!\times\!10^{-21}$ & $  2.0\!\times\!10^{0 }$ & $  3.9\!\times\!10^{2}$ & $ 1.0$ & $ 0$ & $\infty$ & $ 1.0$ & $ 8.0$ & $ 5.0$ & $1/3$ & $\infty$ & $\infty$ \\
07) GT256mM5                       & 256  & mix  & $  3.3\!\times\!10^{-21}$ & $  2.0\!\times\!10^{0 }$ & $  3.9\!\times\!10^{2}$ & $0.99$ & $ 0$ & $\infty$ & $0.98$ & $ 5.4$ & $ 5.0$ & $0.4$ & $\infty$ & $\infty$ \\
08) GT256cM5                       & 256  & comp & $  3.3\!\times\!10^{-21}$ & $  2.0\!\times\!10^{0 }$ & $  3.9\!\times\!10^{2}$ & $0.91$ & $ 0$ & $\infty$ & $0.82$ & $ 1.5$ & $ 4.5$ & $1$   & $\infty$ & $\infty$ \\
\hline
09) GT128sM10                      & 128  & sol  & $  8.2\!\times\!10^{-22}$ & $  8.0\!\times\!10^{0 }$ & $  6.2\!\times\!10^{3}$ & $ 2.1$ & $ 0$ & $\infty$ & $ 1.1$ & $ 11.$ & $ 10.$ & $1/3$ & $\infty$ & $\infty$ \\
10) GT256sM10                      & 256  & sol  & $  8.2\!\times\!10^{-22}$ & $  8.0\!\times\!10^{0 }$ & $  6.2\!\times\!10^{3}$ & $ 2.1$ & $ 0$ & $\infty$ & $ 1.1$ & $ 12.$ & $ 10.$ & $1/3$ & $\infty$ & $\infty$ \\
11) GT512sM10                      & 512  & sol  & $  8.2\!\times\!10^{-22}$ & $  8.0\!\times\!10^{0 }$ & $  6.2\!\times\!10^{3}$ & $ 2.1$ & $ 0$ & $\infty$ & $ 1.1$ & $ 12.$ & $ 10.$ & $1/3$ & $\infty$ & $\infty$ \\
12) GT512mM10\phantom{B10} (s1)    & 512  & mix  & $  8.2\!\times\!10^{-22}$ & $  8.0\!\times\!10^{0 }$ & $  6.2\!\times\!10^{3}$ & $ 2.1$ & $ 0$ & $\infty$ & $ 1.1$ & $ 4.5$ & $ 11.$ & $0.4$ & $\infty$ & $\infty$ \\
13) GT512mM10B1\phantom{0} (s1)    & 512  & mix  & $  8.2\!\times\!10^{-22}$ & $  8.0\!\times\!10^{0 }$ & $  6.2\!\times\!10^{3}$ & $ 2.1$ & $ 1$ & $ 8.2$ & $ 1.1$ & $ 5.4$ & $ 10.$ & $0.4$ & $ 2.8$ & $ 12.$ \\
14) GT512mM10\phantom{B10} (s2)    & 512  & mix  & $  8.2\!\times\!10^{-22}$ & $  8.0\!\times\!10^{0 }$ & $  6.2\!\times\!10^{3}$ & $ 2.2$ & $ 0$ & $\infty$ & $ 1.2$ & $ 8.4$ & $ 11.$ & $0.4$ & $\infty$ & $\infty$ \\
15) GT512mM10B1\phantom{0} (s2)    & 512  & mix  & $  8.2\!\times\!10^{-22}$ & $  8.0\!\times\!10^{0 }$ & $  6.2\!\times\!10^{3}$ & $ 2.2$ & $ 1$ & $ 8.2$ & $ 1.2$ & $ 9.5$ & $ 11.$ & $0.4$ & $ 1.8$ & $ 10.$ \\
16) GT256mM10\phantom{B10} (s3)    & 256  & mix  & $  8.2\!\times\!10^{-22}$ & $  8.0\!\times\!10^{0 }$ & $  6.2\!\times\!10^{3}$ & $ 2.0$ & $ 0$ & $\infty$ & $ 1.0$ & $ 5.9$ & $ 10.$ & $0.4$ & $\infty$ & $\infty$ \\
17) GT512mM10\phantom{B10} (s3)    & 512  & mix  & $  8.2\!\times\!10^{-22}$ & $  8.0\!\times\!10^{0 }$ & $  6.2\!\times\!10^{3}$ & $ 2.0$ & $ 0$ & $\infty$ & $ 1.0$ & $ 5.9$ & $ 10.$ & $0.4$ & $\infty$ & $\infty$ \\
18) GT512mM10B1\phantom{0} (s3)    & 512  & mix  & $  8.2\!\times\!10^{-22}$ & $  8.0\!\times\!10^{0 }$ & $  6.2\!\times\!10^{3}$ & $ 2.0$ & $ 1$ & $ 8.2$ & $0.97$ & $ 6.4$ & $ 9.9$ & $0.4$ & $ 3.6$ & $ 13.$ \\
19) GT256mM10B3\phantom{0} (s3)    & 256  & mix  & $  8.2\!\times\!10^{-22}$ & $  8.0\!\times\!10^{0 }$ & $  6.2\!\times\!10^{3}$ & $ 1.8$ & $ 3$ & $0.92$ & $0.81$ & $ 8.4$ & $ 9.0$ & $0.4$ & $0.20$ & $ 2.9$ \\
20) GT512mM10B3\phantom{0} (s3)    & 512  & mix  & $  8.2\!\times\!10^{-22}$ & $  8.0\!\times\!10^{0 }$ & $  6.2\!\times\!10^{3}$ & $ 1.8$ & $ 3$ & $0.92$ & $0.83$ & $ 8.7$ & $ 9.1$ & $0.4$ & $0.18$ & $ 2.7$ \\
21) GT256mM10B10 (s3)              & 256  & mix  & $  8.2\!\times\!10^{-22}$ & $  8.0\!\times\!10^{0 }$ & $  6.2\!\times\!10^{3}$ & $ 1.8$ & $10$ & $0.08$ & $0.79$ & $ 6.6$ & $ 8.9$ & $0.4$ & $0.04$ & $ 1.3$ \\
22) GT128cM10                      & 128  & comp & $  8.2\!\times\!10^{-22}$ & $  8.0\!\times\!10^{0 }$ & $  6.2\!\times\!10^{3}$ & $ 1.8$ & $ 0$ & $\infty$ & $0.81$ & $ 1.2$ & $ 9.0$ & $1$   & $\infty$ & $\infty$ \\
23) GT256cM10                      & 256  & comp & $  8.2\!\times\!10^{-22}$ & $  8.0\!\times\!10^{0 }$ & $  6.2\!\times\!10^{3}$ & $ 1.8$ & $ 0$ & $\infty$ & $0.85$ & $ 1.1$ & $ 9.2$ & $1$   & $\infty$ & $\infty$ \\
24) GT512cM10                      & 512  & comp & $  8.2\!\times\!10^{-22}$ & $  8.0\!\times\!10^{0 }$ & $  6.2\!\times\!10^{3}$ & $ 1.9$ & $ 0$ & $\infty$ & $0.87$ & $ 1.1$ & $ 9.4$ & $1$   & $\infty$ & $\infty$ \\
\hline
25) GT256sM20                      & 256  & sol  & $  2.1\!\times\!10^{-22}$ & $  3.2\!\times\!10^{1 }$ & $  9.9\!\times\!10^{4}$ & $ 4.1$ & $ 0$ & $\infty$ & $ 1.0$ & $ 11.$ & $ 20.$ & $1/3$ & $\infty$ & $\infty$ \\
26) GT256mM20                      & 256  & mix  & $  2.1\!\times\!10^{-22}$ & $  3.2\!\times\!10^{1 }$ & $  9.9\!\times\!10^{4}$ & $ 4.2$ & $ 0$ & $\infty$ & $ 1.1$ & $ 4.5$ & $ 21.$ & $0.4$ & $\infty$ & $\infty$ \\
27) GT256cM20                      & 256  & comp & $  2.1\!\times\!10^{-22}$ & $  3.2\!\times\!10^{1 }$ & $  9.9\!\times\!10^{4}$ & $ 4.0$ & $ 0$ & $\infty$ & $ 1.0$ & $0.60$ & $ 20.$ & $1$   & $\infty$ & $\infty$ \\
\hline
28) GT256sM50                      & 256  & sol  & $  3.3\!\times\!10^{-23}$ & $  2.0\!\times\!10^{2 }$ & $  3.9\!\times\!10^{6}$ & $ 10.$ & $ 0$ & $\infty$ & $ 1.1$ & $ 12.$ & $ 52.$ & $1/3$ & $\infty$ & $\infty$ \\
29) GT512sM50                      & 512  & sol  & $  3.3\!\times\!10^{-23}$ & $  2.0\!\times\!10^{2 }$ & $  3.9\!\times\!10^{6}$ & $ 10.$ & $ 0$ & $\infty$ & $ 1.1$ & $ 13.$ & $ 52.$ & $1/3$ & $\infty$ & $\infty$ \\
30) GT256mM50                      & 256  & mix  & $  3.3\!\times\!10^{-23}$ & $  2.0\!\times\!10^{2 }$ & $  3.9\!\times\!10^{6}$ & $ 10.$ & $ 0$ & $\infty$ & $ 1.0$ & $ 7.0$ & $ 51.$ & $0.4$ & $\infty$ & $\infty$ \\
31) GT512mM50                      & 512  & mix  & $  3.3\!\times\!10^{-23}$ & $  2.0\!\times\!10^{2 }$ & $  3.9\!\times\!10^{6}$ & $ 10.$ & $ 0$ & $\infty$ & $ 1.1$ & $ 7.4$ & $ 51.$ & $0.4$ & $\infty$ & $\infty$ \\
32) GT256cM50                      & 256  & comp & $  3.3\!\times\!10^{-23}$ & $  2.0\!\times\!10^{2 }$ & $  3.9\!\times\!10^{6}$ & $ 9.8$ & $ 0$ & $\infty$ & $0.95$ & $0.54$ & $ 49.$ & $1$   & $\infty$ & $\infty$ \\
33) GT512cM50                      & 512  & comp & $  3.3\!\times\!10^{-23}$ & $  2.0\!\times\!10^{2 }$ & $  3.9\!\times\!10^{6}$ & $ 9.9$ & $ 0$ & $\infty$ & $0.99$ & $0.56$ & $ 50.$ & $1$   & $\infty$ & $\infty$ \\
34) GT1024cM50                     & 1024 & comp & $  3.3\!\times\!10^{-23}$ & $  2.0\!\times\!10^{2 }$ & $  3.9\!\times\!10^{6}$ & $ 10.$ & $ 0$ & $\infty$ & $1.00$ & $0.55$ & $ 50.$ & $1$   & $\infty$ & $\infty$ \\
\hline
\end{tabular}

\textbf{Notes.} Column (1): simulation name. Columns (2--10): maximum grid resolution $N_\mathrm{res}$ in one direction of the 3D, cubic domain, mode of forcing (solenoidal, mixed, compressive), mean density $\meanrho$, linear box size $L$, total mass $M_c$, velocity dispersion on the box scale $\sigma_V$, mean magnetic field strength $B_0$ (in $z$-direction of the domain), initial plasma $\beta_0$, and virial parameter $\alphavircirc=5\sigma_V^2 L/(6G\mc)$ for a spherical-cloud approximation with uniform density. Columns (11--15): time-averaged virial parameter $\alphavir = 2E_\mathrm{kin}/\left|E_\mathrm{grav}\right|$ of the 3D, inhomogeneous density field in the regime of fully-developed turbulence, sonic Mach number $\mach$, forcing parameter $b$, ratio of thermal to magnetic pressure (plasma $\beta$), and Alfv\'en Mach number $\macha$. To guide the eye, horizontal lines separate models with different sonic Mach number. See Paper I for details of the simulation parameters.
\end{table*}

\subsection{Initial Conditions, Procedures, and List of Models} \label{sec:ics}
We start our numerical experiments with uniform density and zero velocities. The forcing term ${\bf F_\mathrm{stir}}$ (see Section~\ref{sec:forcing}) drives random motions, until a state of fully-developed, supersonic turbulence is reached after two large-scale turbulent crossing times, $2T=L/(\mach\cs)$, as found in previous studies \citep[e.g.,][]{KlessenHeitschMacLow2000,Klessen2001,HeitschMacLowKlessen2001,FederrathKlessenSchmidt2009,FederrathDuvalKlessenSchmidtMacLow2010,PriceFederrath2010,MicicEtAl2012}. This fully-developed turbulent state is the initial condition for our star-formation experiments, when self-gravity is added and formation of sink particles is allowed. We denote this time as $t=0$ in the following.

All our numerical simulations and their basic parameters are listed in Table~\ref{tab:sims}. These parameters were chosen to roughly follow observed properties of molecular clouds, covering a range of cloud sizes $L\sim0.3$--$200\,\pc$, masses $M_c\sim300$ to $4\times10^6\,\msol$, and velocity dispersions $\sigma_V\sim0.6$--$10\,\km\,\s^{-1}$ \citep[e.g.,][]{Larson1981,SolomonEtAl1987,FalgaronePugetPerault1992,OssenkopfMacLow2002,HeyerBrunt2004,HeyerEtAl2009,RomanDuvalEtAl2011}, with typical cloud scalings summarized and discussed in \citet{MacLowKlessen2004} and \citet{McKeeOstriker2007}. For the MHD models, we chose magnetic field strengths consistent with the range observed in clouds of that size and density \citep[see][]{Crutcher1999,HeilesTroland2005,CrutcherEtAl2010}. We compare models with initial line-of-sight magnetic field strengths $B_{0,z} = 1$, 3, and $10\,\mu\Gauss$.

\subsection{Definition of the SFE}
We use the standard definition of the SFE \citep[e.g.,][]{MyersEtAl1986}, which is the mass in star-forming gas (stars), $\mstar$, divided by the total mass of the cloud,
\begin{equation} \label{eq:sfe}
\sfe \equiv \frac{\mstar}{\mstar+\mgas} = \frac{\mstar}{M_c}.
\end{equation}
We measure $\mstar$ in the simulations simply by taking the sum of all sink particle masses formed in the simulation at a given time. The denominator in Equation~(\ref{eq:sfe}) is constant over time and given by the total cloud mass, $M_c$, in Table~\ref{tab:sims}, because our computational boxes are closed, without global gas inflow or outflow. Gas that is turned into sinks is removed from the gas phase to conserve the total mass in the system.

\begin{figure*}[t]
\centerline{
\includegraphics[width=0.95\linewidth]{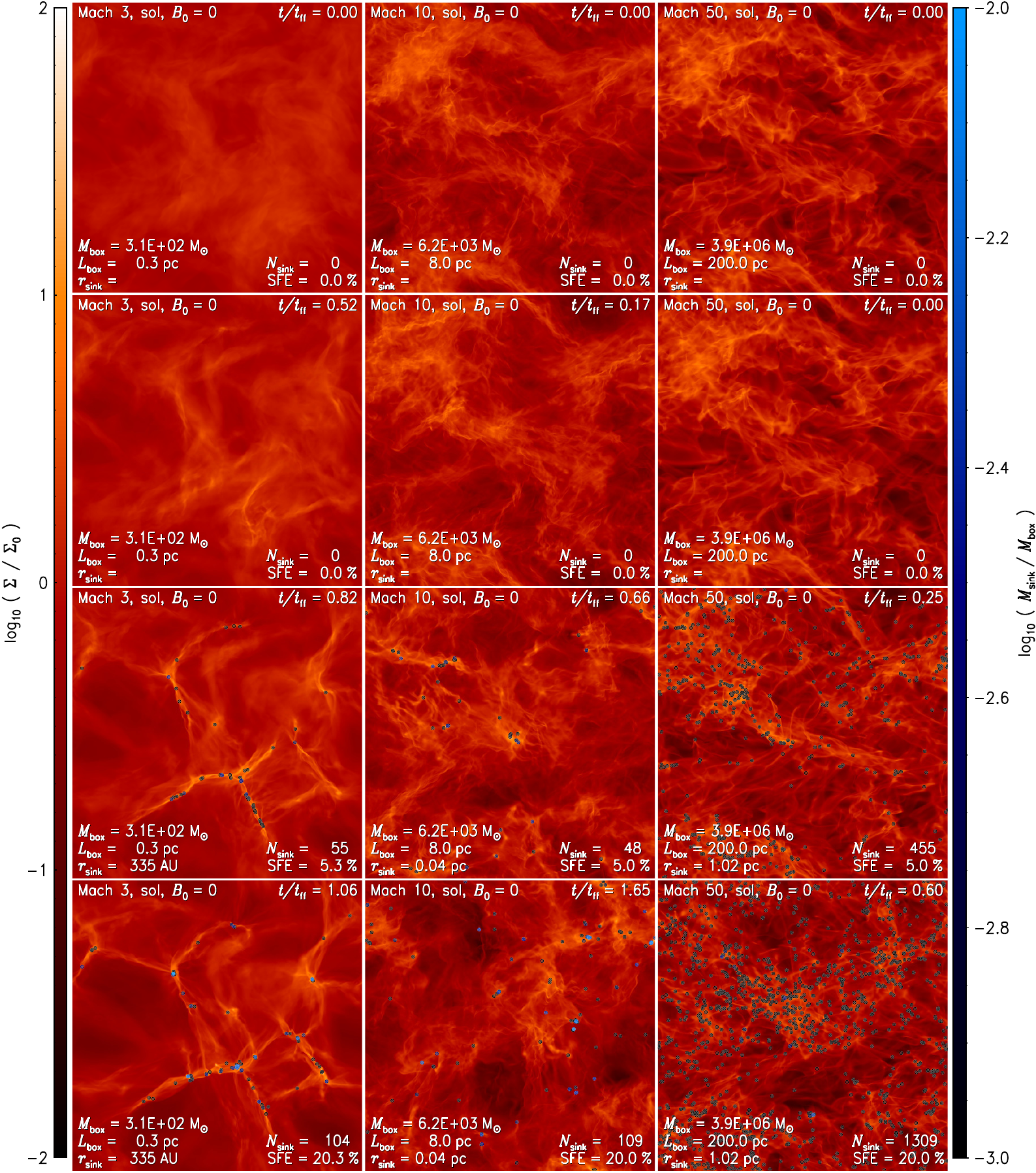}
}
\caption{Time evolution of column density (normalized to the mean column density $\Sigma_0$) for simulations with solenoidal forcing at $512^3$ resolution and $\mach\sim3$ (\emph{left}), $\mach\sim10$ (\emph{middle}), and $\mach\sim50$ (\emph{right}), when $t=0$ (state of fully developed turbulence; \emph{top row}), and when $\sfe=0$ (right before the first sink particle forms; \emph{2.~row}), $\sfe=5\%$ (\emph{3.~row}), and $\sfe=20\%$ (\emph{bottom row}). Sink particles are shown as circles with radius $\rsink$. Simulation parameters and variables are indicated in each panel and in Table~\ref{tab:sims}.}
\label{fig:imagessol}
\end{figure*}

\begin{figure*}[t]
\centerline{
\includegraphics[width=0.95\linewidth]{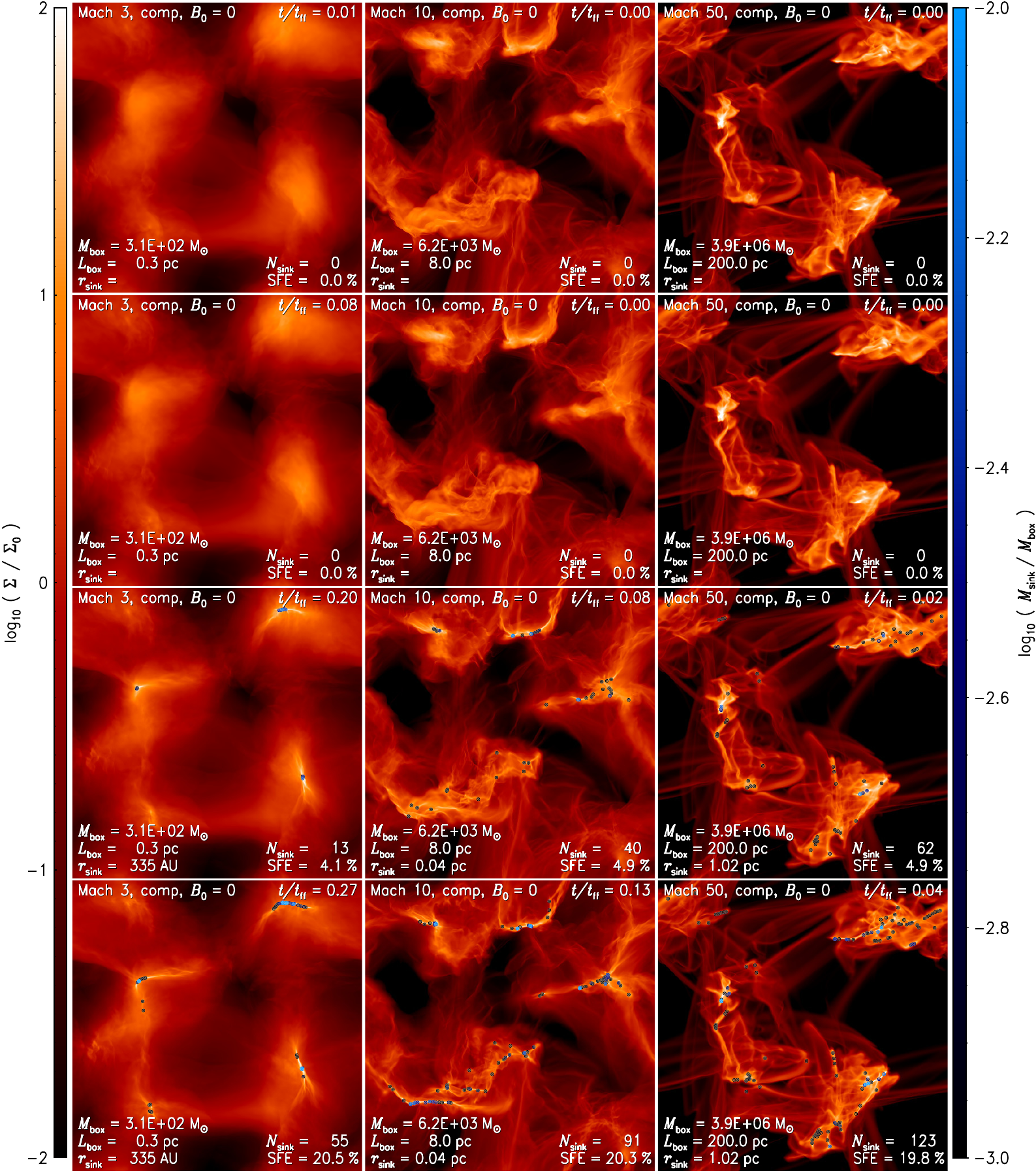}
}
\caption{Same as Figure~\ref{fig:imagessol}, but for compressive forcing of the turbulence.}
\label{fig:imagescomp}
\end{figure*}

\section{Evolution of the Cloud Structure} \label{sec:morphology}

After the initial turbulent state has been established (see Section~\ref{sec:ics}), we study the subsequent evolution under the influence of self-gravity. First, we look at how the column density structure changes with increasing $\sfe$ in Figures~\ref{fig:imagessol} and~\ref{fig:imagescomp}. We see that density contrasts increase with increasing sonic Mach number (from left to right: $\mach\sim3$, 10, and 50). For a given Mach number, solenoidal forcing (Figure~\ref{fig:imagessol}) produces significantly less variation in the column density than compressive forcing (Figure~\ref{fig:imagescomp}). Star formation (increasing $\sfe$ from top to bottom in the figures) affects mostly the small scale structure where collapse proceeds, but also reaches to large scales, best seen in the runs with $\mach\sim3$. There, we also find evidence for global collapse, while star formation is locally much faster in the high-Mach number ($\mach\sim10$ and 50) cases, such that no significant global collapse occurs even when $\sfe=20\%$ is reached. This is because in the high-Mach number cases, turbulence alone provides significant local compression, such that small-scale structures quickly proceed to collapse. For instance, in the most extreme case ($\mach\sim50$ with compressive forcing; right panels in Figure~\ref{fig:imagescomp}), the large-scale structure does not change at all during the short time required to wind up 20\% of the gas in bound cores and stars. In contrast, the turbulent density seeds in the $\mach\sim3$ runs have such a low amplitude that large-scale gravitational contraction is required to increase the local densities to the point when star formation can proceed. This behavior is reflected in the SFRs of all our simulations, ranging over two orders of magnitude, depending on the virial parameter, the turbulent forcing, the sonic Mach number, and the Alfv\'en Mach number, studied in detail in Paper I.

The column density evolution and dependence on the forcing and the sonic Mach number give us a visual impression of the time dependence and spatial distribution of star formation. We immediately see that the density variance depends on the sonic Mach number and the forcing already in the purely turbulent regime (top rows in Figures~\ref{fig:imagessol} and~\ref{fig:imagescomp}) and become stronger on small scales, where star formation occurs. Stars form primarily in dense filaments, with more massive stars being located mostly at intersections of multiple filaments or where filaments seem to bend. There, mass is transported to the protostars more efficiently, because it is funneled through multiple filaments toward their intersections where gravitational focusing occurs \citep{BurkertHartmann2004}. Similar structures and the correlation of star formation with filaments and their intersections have been observed in molecular clouds \citep[e.g.,][]{AndreEtAl2010,SchneiderEtAl2010,ArzoumanianEtAl2011,SchneiderEtAl2012}. Supersonic turbulence and gravity are the natural physical processes that can produce such filaments in gas that can cool sufficiently to roughly maintain a constant temperature over a wide range of densities \citep[see][and references therein]{PetersEtAl2012}, as seen in the observations and in the present simulations.

In order to gain more insight into the correlation of gas density with star formation, we investigate density and column density PDFs in the following.

\section{The Density PDF} \label{sec:pdfs}

The density PDF of the turbulent gas is a key for analytic models of the core and stellar initial mass functions \citep{PadoanNordlund2002,HennebelleChabrier2008,HennebelleChabrier2009,Elmegreen2011,VeltchevKlessenClark2011,DonkovVeltchevKlessen2012,ParravanoSanchezAlfaro2012,Hopkins2012b,Hopkins2012c}, the Kennicutt-Schmidt relation \citep{KrumholzMcKee2005,Tassis2007}, the SFE \citep{Elmegreen2008}, and the SFR \citep[][Paper I]{KrumholzMcKee2005,PadoanNordlund2011,HennebelleChabrier2011}. We investigate the dependences of the PDF on the forcing of the turbulence, the Mach number, and the magnetic field, and pay specific attention to its dependence on the SFE in the following.

\subsection{The Density PDF of Supersonic, Isothermal Turbulence}

Studies of non-gravitating, isothermal turbulence show that the density PDF is close to a log-normal distribution,
\begin{equation}
p_s(s)=\frac{1}{\sqrt{2\pi\sigs^2}}\exp\left(-\frac{(s-s_0)^2}{2\sigs^2}\right)\,,
\label{eq:pdf}
\end{equation}
expressed in terms of the logarithmic density,
\begin{equation} \label{eq:s}
s\equiv\ln{(\rho/\meanrho)}\,.
\end{equation}
The PDF is a normal (Gaussian) distribution in $s$, and hence a log-normal distribution in $\rho$. The quantities $\meanrho$ and $\means$ denote the mean density and mean logarithmic density, the latter of which is related to the standard deviation $\sigs$ by $\means=-\sigs^2/2$ due to the normalization and mass-conservation constraints of the PDF \citep{Vazquez1994,LiKlessenMacLow2003,FederrathKlessenSchmidt2008}.

The standard deviation $\sigs$ in Equation~(\ref{eq:pdf}) is a measure of how much the density varies in a turbulent medium and depends on 1) the amount of compression induced by the turbulent forcing mechanism, 2) the sonic Mach number, and 3) the magnetic pressure. \citet{MolinaEtAl2012} provide a rigorous derivation of the variance of the PDF,
\begin{equation}
\sigs^2=\ln\left(1+b^2\mach^2\frac{\beta}{\beta+1}\right)\,,
\label{eq:sigs}
\end{equation}
with the forcing parameter $b$, the sonic RMS Mach number $\mach$, and the ratio of thermal to magnetic pressure, which can be expressed as the ratio of sound to Alfv\'en speed or as the ratio of Alfv\'en to sonic Mach number, $\beta=2\cs^2/\va^2=2\macha^2/\mach^2$. The forcing parameter $b$ in Equation~(\ref{eq:sigs}) was shown to vary smoothly between $b\sim1/3$ for purely solenoidal (divergence-free) forcing, and $b\sim1$ for purely compressive (curly-free) forcing of the turbulence \citep[see Section~\ref{sec:forcing};][]{FederrathKlessenSchmidt2008,SchmidtEtAl2009,FederrathDuvalKlessenSchmidtMacLow2010,SeifriedSchmidtNiemeyer2011,MicicEtAl2012,KonstandinEtAl2012}. A stochastic mixture of forcing modes in 3D space leads to $b\sim0.4$ \citep[see Figure~8 in][]{FederrathDuvalKlessenSchmidtMacLow2010}.

When gravity is included and significant collapse sets in, the density PDF develops a power-law tail at high densities \citep[][]{Klessen2000}, which we concentrate on in the following.

\subsection{The Density PDF of Self-gravitating Turbulence}

\begin{figure*}[t]
\centerline{
\includegraphics[width=0.94\linewidth]{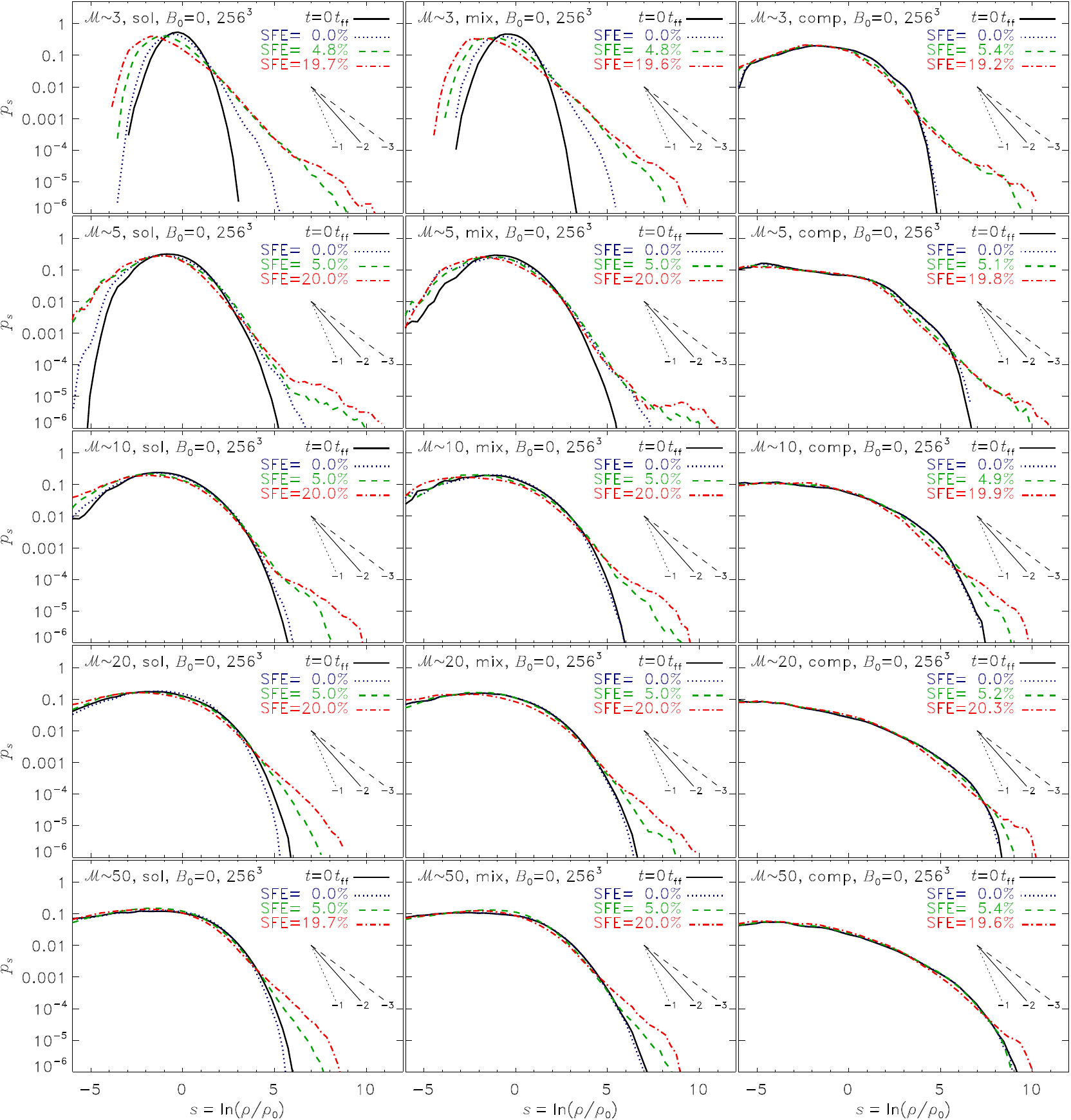}
}
\caption{Probability distribution functions (PDFs) of the logarithmic density $s\equiv\ln(\rho/\meanrho)$ for hydrodynamic ($B_0=0$) models with solenoidal (\emph{left}), mixed (\emph{middle}), and compressive forcing (\emph{right}) and sonic Mach numbers $\mach\sim3$, 5, 10, 20, 50 (\emph{from top to bottom}) at different times: $t=0$: state of fully developed turbulence (solid), $\sfe=0$: right before first sink particle forms (dotted), $\sfe\sim5\%$ (dashed), and $\sfe\sim20\%$ (dot-dashed). The standard deviation of the PDFs increases with increasing Mach number and more compressive forcing. All PDFs develop power-law tails at high densities of flattening slope with increasing $\sfe$. The three power-law lines below the legend in each panel show slopes of power-law PDFs equivalent to PDFs obtained for volumetric, radial-density profiles $\rho(r)\propto r^{-\kappa}$ with $\kappa=1$ (dotted), $\kappa=2$ (solid), and $\kappa=3$ (dashed), according to Equation~(\ref{eq:spowerlaw}).}
\label{fig:pdfs}
\end{figure*}

\begin{figure*}[t]
\centerline{
\includegraphics[width=0.94\linewidth]{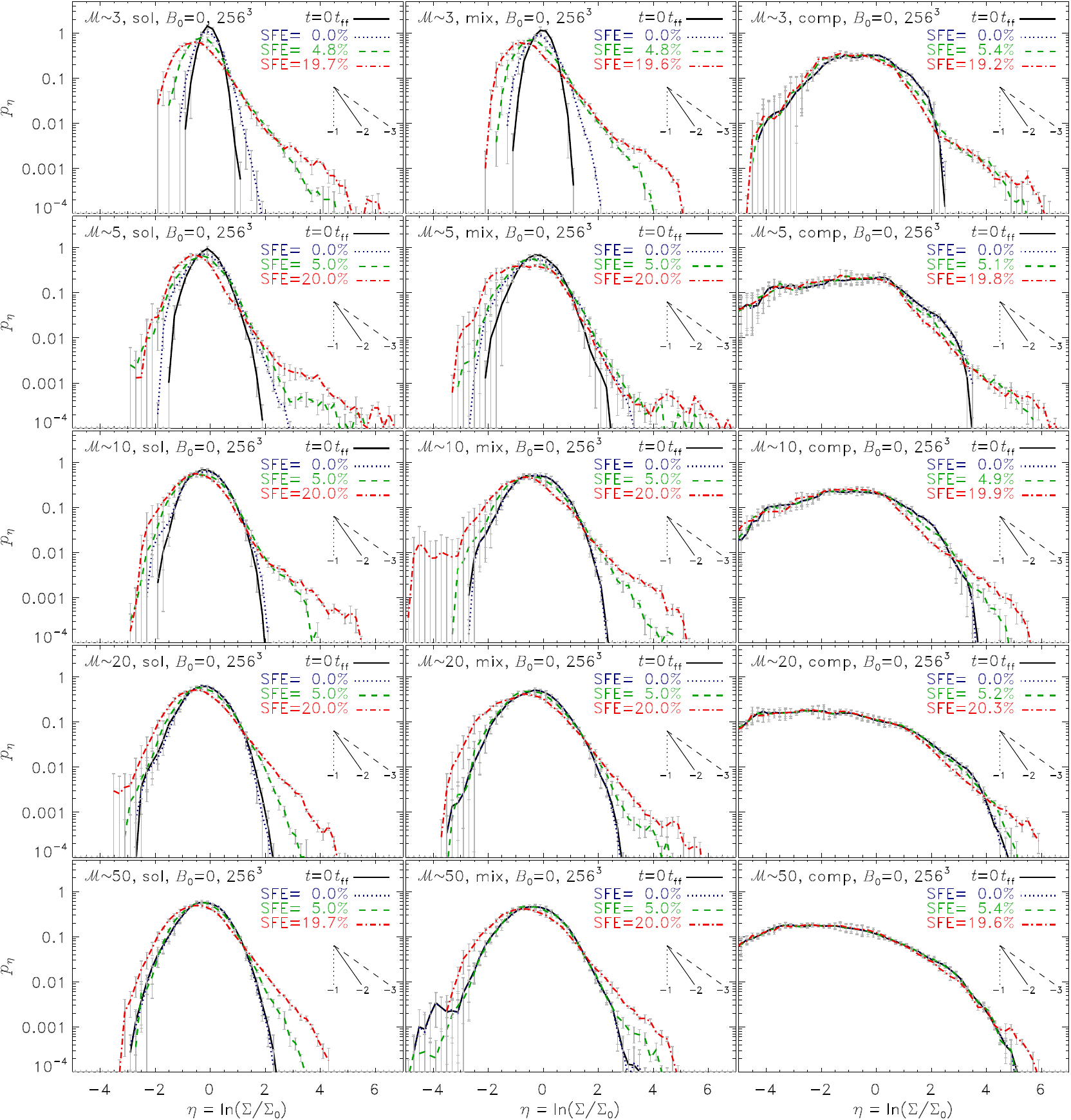}
}
\caption{Same as Figure~\ref{fig:pdfs}, but for the logarithmic column density $\eta\equiv\ln(\Sigma/\Sigma_0)$ with the mean column density $\Sigma_0$. Error bars indicate the uncertainty in the PDFs when three different projection directions (along the $x$, $y$, or $z$ axes) of the same cloud are compared. The column density PDFs are more noisy, but follow the same trends as the volumetric density PDFs in Figure~\ref{fig:pdfs}. However, the standard deviation is smaller in $p_\eta$ than $p_s$, because of projection \citep{FischeraDopita2004,FederrathDuvalKlessenSchmidtMacLow2010,BruntFederrathPrice2010a,BruntFederrathPrice2010b,Seon2012}, and the power-law tails have a different slope (compare Equations~\ref{eq:spowerlaw} and~\ref{eq:etapowerlaw}).}
\label{fig:coldenspdfs}
\end{figure*}

\begin{figure*}[t]
\centerline{
\includegraphics[width=0.94\linewidth]{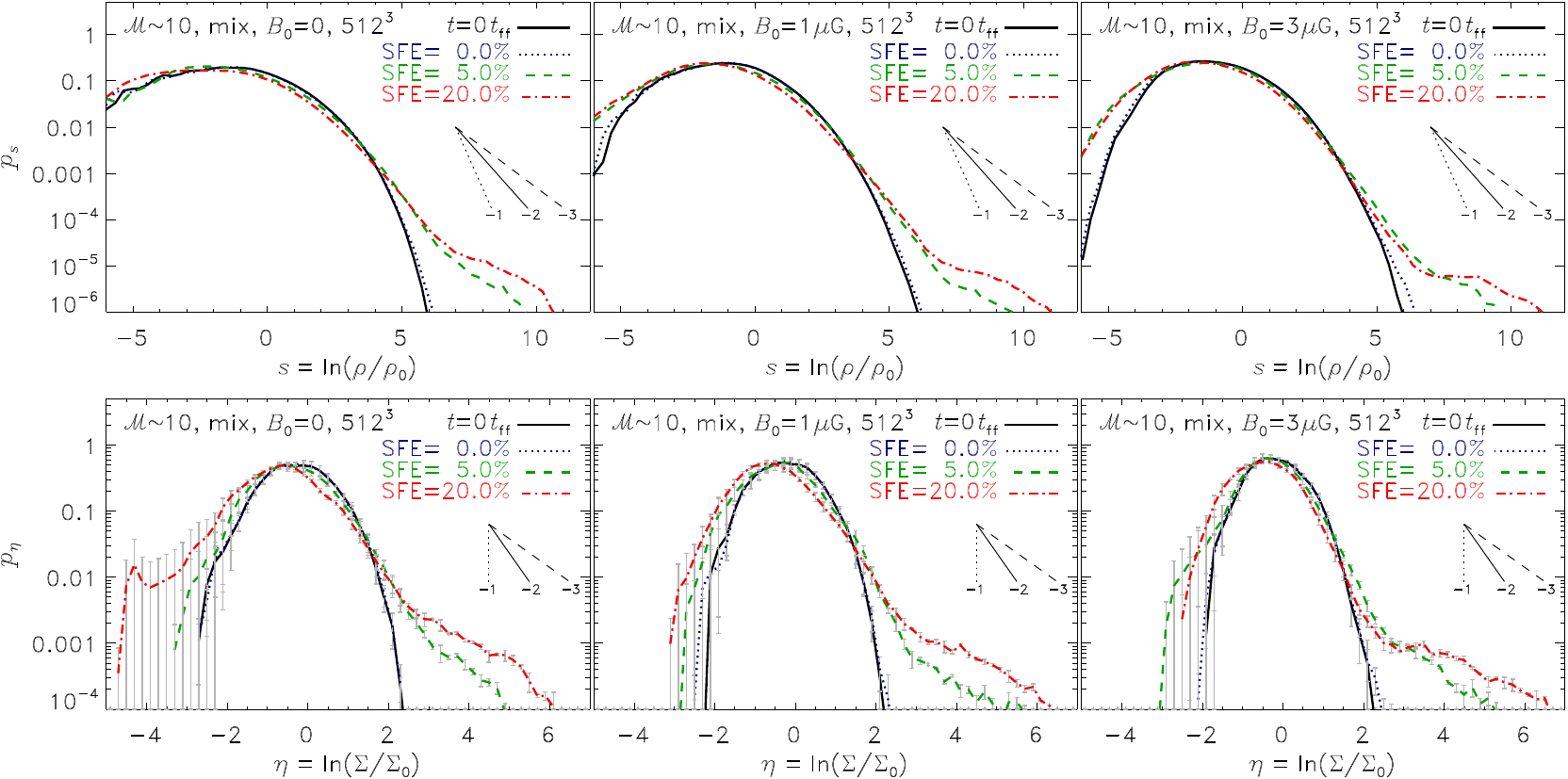}
}
\caption{\emph{Top row}: same as Figure~\ref{fig:pdfs}, but for models with mixed-forcing at $\mach\sim10$ and varying initial magnetic field $B_0=0$ \emph{(left}), $1\,\mu\Gauss$ (\emph{middle}), and $3\,\mu\Gauss$ (\emph{right}) at $512^3$ resolution. \emph{Bottom row}: same as top row, but for the column density PDFs as in Figure~\ref{fig:coldenspdfs}. The density variance decreases with increasing magnetic field strength, in agreement with the theoretical prediction given by Equation~(\ref{eq:sigs}). High-density power-law tails develop in all models, but no clear trend with magnetic field strength is seen.}
\label{fig:magpdfs}
\end{figure*}

\subsubsection{Volumetric Density PDFs}

Figure~\ref{fig:pdfs} shows the volumetric density PDFs of $s=\ln(\rho/\meanrho)$ for hydrodynamic runs with different forcing (from left to right: solenoidal, mixed, and compressive forcing), and different Mach number (from top to bottom: $\mach\sim3$, 5, 10, 20, and 50) at a fixed resolution of $256^3$ grid cells. Each panel shows the time evolution, with the initial turbulent state $t=0$, with $\sfe=0$, representing the time right before the first sink particle forms, and when the simulation reached $\sfe=5\%$ and 20\%. The initial turbulent state, $t=0$, can be approximated with log-normal distributions, Equation~(\ref{eq:pdf}), in the low-Mach number cases, while larger Mach numbers lead to stronger deviations from the perfect log-normal shape. This is partly caused by numerical and partly caused by physical effects. The numerical effects are poor sampling and limited resolution at high Mach number \citep{PriceFederrath2010,PriceFederrathBrunt2011,KonstandinEtAl2012ApJ}. We explore the resolution dependence and the influence of different random sampling in Appendices~\ref{app:pdfres} and~\ref{app:seeds}. The physical effect introducing deviations from log-normal statistics is an increased intermittency for high-Mach number turbulence, producing larger skewness and kurtosis in the distributions \citep{Klessen2000,KritsukEtAl2007,BurkhartEtAl2009,FederrathDuvalKlessenSchmidtMacLow2010,Hopkins2012d}. Although both numerics and physics affect the wings of the PDFs, we can draw general conclusions from the trends seen in Figure~\ref{fig:pdfs}.

At late times, all simulations develop power-law tails when the gas collapses to high densities. This effect is clearly time-dependent, such that the slope of the tail becomes flatter as more and more gas collapses. This was seen first in smoothed particle hydrodynamics (SPH) simulations by \citet{Klessen2000} and later confirmed in grid simulations with tracer particles \citep{FederrathGloverKlessenSchmidt2008} and in pure grid simulations \citep{DibBurkert2005,VazquezSemadeniEtAl2008,ChoKim2011,KritsukNormanWagner2011,BallesterosEtAl2011,CollinsEtAl2012,SafranekShraderEtAl2012}, though with a significantly narrower parameter range than studied here. All those previous numerical models \citep[except the study by][which on the other hand had a very low resolution of only about $70^3$ effective SPH particles]{Klessen2000} were limited in that they could not follow the collapse to late stages with SFEs much larger than $\sfe\gtrsim0$, because they did not include sink particles (see Section~\ref{sec:sinks}). For instance, \citet{KritsukNormanWagner2011} and \citet{CollinsEtAl2012} had to stop their simulations when the first over-densities started to collapse, because the simulations become prohibitively slow and the resolution criterion for collapsing gas is violated at late times (see Section~\ref{sec:sinks}). Using sink particles here, we can follow the development of the high-density tail in the PDFs to much later times. To guide the eye, we draw power-law PDFs that would be obtained for purely radial density distributions,
\begin{equation} \label{eq:raddens}
\rho(r)\propto r^{-\kappa}\,,
\end{equation}
with $\kappa=1$, 2, and 3 (dotted, solid, and dashed power-law lines in each panel). Here we compare PDF slopes in terms of this equivalent radial density profile. This enables a direct comparison of the slopes in the tails of the density PDF from different numerical, theoretical, and observational studies, because the slope of the equivalent radial profile can always be computed from any representation of the density PDF (e.g., volume-weighted or mass-weighted, linear or logarithmic density, volume or column density). We provide a derivation of the relationship between volume and column density PDFs and their equivalent radial density profiles below. A comparison of high-density PDF tails in terms of equivalent radial density profiles is furthermore motivated by the fact that the dense parts of molecular clouds eventually develop self-gravitating cores, which can be approximated with such radial power-law profiles as studied in theory \citep[e.g.,][]{Larson1969,Penston1969,Shu1977,WhitworthSummers1985}, numerical simulations \citep[e.g.,][]{FosterChevalier1993,BanerjeePudritz2007,FederrathSurSchleicherBanerjeeKlessen2011}, and observations \citep{KoenyvesEtAl2010,ArzoumanianEtAl2011,HillEtAl2011,SchneiderEtAl2012}.

Here we use volume-weighted PDFs of the natural logarithm of density, $p_s$ with $s=\ln(\rho/\meanrho)$, as in Equation~(\ref{eq:pdf}). The PDF $p_s$ is related to the PDF of the density, $p_\rho$ by \citep[e.g.,][]{LiKlessenMacLow2003,FederrathSurSchleicherBanerjeeKlessen2011}
\begin{equation} \label{eq:prho2ps}
p_s = \rho\,p_\rho\,.
\end{equation}
The PDF $p_\rho$ in turn is a measure of the volume fraction with gas of density $\rho$, and is thus defined as
\begin{equation}
p_\rho \propto \deriv V / \deriv\rho = (\deriv V / \deriv r) (\deriv r / \deriv\rho) \propto \rho^{-(3/\kappa+1)}
\end{equation}
with the volume $V\propto r^3$. Substituting this into Equation~(\ref{eq:prho2ps}), we find
\begin{equation} \label{eq:spowerlaw}
p_s \propto \rho^{-3/\kappa} \propto \exp(-3s/\kappa)\,.
\end{equation}
In analogy, we can derive the power-law scaling of the PDF of column density $\Sigma\propto\rho\,r\propto r^{-\kappa+1}$,
\begin{equation}
p_\Sigma \propto \deriv A / \deriv\Sigma \propto \Sigma^{(1+\kappa)/(1-\kappa)}
\end{equation}
with the area $A\propto r^2$. The logarithmic column density PDF of $\eta\equiv\ln(\Sigma/\Sigma_0)$, defined in analogy to $s$, is
\begin{equation} \label{eq:etapowerlaw}
p_\eta = \Sigma\,p_\Sigma \propto \Sigma^{2/(1-\kappa)} \propto \exp\left[2\eta/(1-\kappa)\right]\,,
\end{equation}
relating the slope of the column density PDF to the slope $\kappa$ of the equivalent radial density profile given by Equation~(\ref{eq:raddens}). Note that the column density PDF slope diverges for $\kappa=1$, because the column density, $\Sigma\propto r^{-\kappa+1}$ is constant in that case, which means that $p_\Sigma$ and $p_\eta$ are delta functions.

\subsubsection{Column Density PDFs}

Figure~\ref{fig:coldenspdfs} shows the same as Figure~\ref{fig:pdfs}, but for the logarithmic column density variable $\eta\equiv\ln(\Sigma/\Sigma_0)$, where $\Sigma_0$ is the mean column density of the region considered. The power-law slopes for radial, volumetric density profiles are given in analogy to Figure~\ref{fig:pdfs} for $\kappa=1$, 2, and 3 in each panel. The transformation of slopes between the volumetric and the column density PDFs is obtained by comparing Equations~(\ref{eq:spowerlaw}) and~(\ref{eq:etapowerlaw}). Intermittency and sampling effects are more pronounced in the column density PDFs than in the volumetric density PDFs. We also added error bars to indicate the variation of the $\eta$-PDF when three different projections of the same cloud are considered. The general trends with different models are similar to the volumetric density PDFs of Figure~\ref{fig:pdfs}. The statistics at low column density is more strongly affected by sampling and projection effects, but the power-law tails at high densities are significant, and can be measured in molecular cloud observations \citep[e.g.,][]{KainulainenEtAl2009,SchneiderEtAl2012}.

The power-law slope of the high-density PDF tails depends on the range chosen to fit and on the $\sfe$. It may also depend on physical parameters (different forcing and/or Mach number, etc.), but---unlike the standard deviation of the PDF, Equation~(\ref{eq:sigs})---the power-law slope does not seem to depend strongly on the forcing or the sonic Mach number of the turbulence. Here, we only highlight general trends without fitting slopes, because of limited resolution, sampling, and general uncertainties in setting the fit range. However, we clearly find that the high-density tail in both volumetric and column density PDFs of Figures~\ref{fig:pdfs} and~\ref{fig:coldenspdfs} flattens with increasing $\sfe$, or equivalently, the corresponding radial density profile, Equation~(\ref{eq:raddens}), steepens from $\kappa\lesssim1.5$ for $\sfe\sim0$ to \mbox{$\kappa\sim1.5$--$2.5$} between $\sfe\sim0$ and $\sfe\sim20\%$ in all numerical models. Our estimates of $\kappa$ agree well with the range of equivalent radial power-law slopes inferred in recent \emph{Herschel} observations, \mbox{$\kappa\sim1.5$--$2.5$} \citep{ArzoumanianEtAl2011}, suggesting SFEs in the range $\sfe=0\%$ to $20\%$.

\citet{KritsukNormanWagner2011} provide a very detailed measurement of the slope in their single model with mixed forcing at $\mach\sim6$. They find a global PDF slope corresponding to \mbox{$\kappa\sim1.8$} at the latest time available in their simulation. Steeper global power-law slopes were not obtained in \citet{KritsukNormanWagner2011}, probably because they had to stop the simulation as soon as the first over-density went into collapse. We find a similar global slope in our model with mixed forcing at $\mach\sim5$ when $\sfe\sim0$. However, our results for different forcing and sonic Mach numbers suggest that the slope of the high-density tail of the PDF is not universal. Moreover, the PDF slope is clearly time-dependent and flattens in all our models, consistent with previous studies probing a more limited physical parameter range \citep{ChoKim2011,CollinsEtAl2012}.



\subsubsection{The PDF of Magnetized Clouds}

Figure~\ref{fig:magpdfs} shows a comparison of volumetric (top) and column density PDFs (bottom) for simulations with magnetic fields, $B_0=0$ (left), $1\,\mu\Gauss$ (middle), and $3\,\mu\Gauss$ (right). According to Equation~(\ref{eq:sigs}), the density variance should decrease with increasing magnetic pressure. Indeed, the PDFs become narrower with decreasing plasma $\beta$ or decreasing $\macha$ (see Table~\ref{tab:sims}), as observed in \citet{CollinsEtAl2012}, and quantified in \citet{MolinaEtAl2012}. The high-density power-law slope, however, does not seem to exhibit a clear trend with magnetic field strength. \citet{CollinsEtAl2012} report a weak steepening of the tail with decreasing $\beta$, but that depends again on the time and fit range chosen.

Correlating model parameters with measured PDF slopes for different $\sfe$ would be a natural way to proceed in follow-up studies, in order to learn about the state of a cloud by measuring its column density PDF. For instance, the standard deviation (width) of the PDF shows a clear dependence on the forcing, the Mach number, and the magnetic field strength. These dependences have been studied theoretically and numerically for volumetric density PDFs \citep{PassotVazquez1998,NordlundPadoan1999,FederrathKlessenSchmidt2008,PriceFederrathBrunt2011,MolinaEtAl2012,KonstandinEtAl2012ApJ} and were extended to the study of column density PDFs \citep{FischeraDopita2004,BruntFederrathPrice2010a,BruntFederrathPrice2010b,Brunt2010,BurkhartLazarian2012,KainulainenTan2012,Seon2012}. We clearly see in Figures~\ref{fig:pdfs} and~\ref{fig:coldenspdfs} that the standard deviation of the PDF becomes larger as $\mach$ is increased and as the forcing becomes more compressive, while we see in Figure~\ref{fig:magpdfs} that the standard deviation decreases with increasing magnetic field strength. Thus, a promising goal would be to relate PDF parameters with cloud properties, in order to constrain the evolutionary stage and physical parameters of a cloud, e.g., star-forming versus quiescent \citep{KainulainenEtAl2009}, pressure-confined or not \citep{KainulainenEtAl2011}, influenced by ionizing radiation, or other environmental effects \citep{SchneiderEtAl2012}, by measuring column density PDFs and comparing them with simulations.

\section{Velocity, Density, and Column Density Scalings of Molecular Clouds} \label{sec:scaling}

In this section, we study the spatial scaling of turbulent velocity, density, and column density. In particular, we find that the column density power spectrum is a powerful tool to distinguish star-forming from non-star-forming clouds.

\subsection{Definition of the Fourier Power Spectrum} \label{sec:spectdef}
Fourier power spectra are commonly used in turbulence analysis to study the scaling and correlation of a given physical quantity $q(\ell)$ (e.g., velocity, density, or combinations) with spatial scale $\ell$. Since the analysis is carried out in Fourier space, the spatial scale simply transforms to a wavenumber scale $k=2\pi/\ell$. The $D$-dimensional Fourier transform of $q(\boldsymbol{\ell})$ with $\boldsymbol{\ell}=\{\ell_1,...,\ell_D\}$ is defined as
\begin{equation} \label{eq:ft}
\widehat{q}(\boldsymbol{k}) = \frac{1}{(2\pi L)^{D/2}} \int q(\boldsymbol{\ell})\,e^{-i\,\boldsymbol{k}\cdot\boldsymbol{\ell}}\,\deriv^D \ell\,,
\end{equation}
where we denote the Fourier transform of $q(\boldsymbol{\ell})$ with $\widehat{q}(\boldsymbol{k})$, where $\boldsymbol{\ell},\,\boldsymbol{k} \in \Re^D$. The spatial scale $\ell$ is in the range $[0,L]$ with the maximum spatial scale $L$ corresponding to the smallest wavenumber $2\pi/L$, i.e., the largest scale of a cloud or in a column density map. Note that in this definition of the $D$-dimensional Fourier transform, $\widehat{q}$ has units of $[qL^{D/2}]=[qk^{-D/2}]$, such that $\int \widehat{q}\cdot\widehat{q}^\star \deriv^D k$ always yields a quantity with units of the original physical variable $q^2$.

With the definition of the Fourier transform, Equation~(\ref{eq:ft}), the 3D ($D=3$) Fourier power spectrum of $q$ is given by
\begin{equation} \label{eq:pthreed}
P_\mathrm{3D}(q,k) = \langle \widehat{q}\cdot\widehat{q}^\star\,4\pi k^2 \rangle_k\,,
\end{equation}
as an average of $\widehat{q}\cdot\widehat{q}^\star$ over a spherical shell with radius $k=|\boldsymbol{k}|$ and thickness $\deriv k$ in Fourier space, where $\widehat{q}^\star$ denotes the complex conjugate of $\widehat{q}$. In analogy, the two-dimensional (2D, $D=2$) power spectrum,
\begin{equation} \label{eq:ptwod}
P_\mathrm{2D}(q,k) = \langle \widehat{q}\cdot\widehat{q}^\star\,2\pi k \rangle_k\,,
\end{equation}
is defined as an average over a ring with radius $k$ and thickness $\deriv k$. As a result of the definition of the $D$-dimensional Fourier transform, both $P_\mathrm{3D}(q,k)$ and $P_\mathrm{2D}(q,k)$ have the same units, $[q^2/k]$.

\subsection{Definition of the $\Delta$-variance Spectrum} \label{sec:dvardef}
The $\Delta$-variance technique provides a complementary method for measuring the scaling exponent of Fourier spectra in the physical domain by applying a wavelet transformation \citep{StutzkiEtAl1998}. We here use the $\Delta$-variance tool developed and provided by \citet{OssenkopfKripsStutzki2008a} for 2D maps. This tool implements an improved version of the original $\Delta$-variance \citep{StutzkiEtAl1998,BenschStutzkiOssenkopf2001}, capable of treating non-periodic edges and taking into account variations in signal-to-noise, which is important for applying the method to observational maps \citep[e.g.,][]{SchneiderEtAl2011}. The $\Delta$-variance measures the amount of structure in an observational map, by filtering the dataset $q(\boldsymbol{\ell})$ with a symmetric up-down-function $\bigodot\!_\ell$ (typically a French-hat or Mexican-hat filter) of size $\ell$, and computing the variance of the filtered dataset as a function of filter size $\ell$. The $\Delta$-variance is defined as
\begin{equation} \label{eq:dvardef}
\sigma_\Delta^2(q,\ell) = \left<\left(q(\vect{x})\ast\bigodot\!\frac{\!}{\!}_\ell(\vect{x})\right)^2\right>_{\!\vect{x}}\,,
\end{equation}
where the average is computed over all data points at positions $\vect{x}=\{x_1,x_2\}$ for a 2D map $q(\vect{x})$. The operator $\ast$ denotes a convolution, which in practice, is carried out in Fourier space. We use the original French-hat filter with a diameter ratio of $3.0$ as in previous studies using the $\Delta$-variance technique \citep[e.g.,][]{StutzkiEtAl1998,MacLowOssenkopf2000,OssenkopfKlessenHeitsch2001,OssenkopfMacLow2002,OssenkopfEtAl2006}. There is some uncertainty introduced by the filter type and diameter chosen, but for turbulent systems, the $\Delta$-variance does not significantly depend on that choice \citep{OssenkopfKripsStutzki2008a}. Here, we apply the $\Delta$-variance only to the column density, i.e., $q\equiv\Sigma$ in Equation~(\ref{eq:dvardef}). The relation between $\dvar(\Sigma,\ell)$ and the Fourier power spectrum of the density, $\pthreed(\rho,k)$, is explained in the next section.

\subsection{Relation between $\pthreed$ and $\dvar$}
Assuming a power-law scaling of the 3D density power spectrum,
\begin{equation} \label{eq:alpha}
\pthreed(\rho,k)\propto k^\alpha
\end{equation}
and a power-law scaling of the 2D column density power spectrum
\begin{equation}
\ptwod(\Sigma,k)\propto k^\beta\,,
\end{equation}
we see from the definitions~(\ref{eq:pthreed}) and~(\ref{eq:ptwod}) that they are directly related by
\begin{equation} \label{eq:ptwodtothreedrelation}
\pthreed=2 k \ptwod \propto k^\alpha \propto k^{\beta+1}\,.
\end{equation}
Moreover, considering the spatial scaling of the column density $\Sigma$ in real space (instead of wavenumber space), we have to recall that $\ptwod(\Sigma)\propto \deriv\Sigma^2/\deriv k$. Thus, 
\begin{equation}
\Sigma^2\propto \ptwod k \propto \pthreed \propto k^{\beta+1} \propto \ell^{-(\beta+1)} \propto \ell^{-\alpha}\,.
\end{equation}
This result shows that the $\Delta$-variance power spectrum of $\Sigma$ defined in Equation~(\ref{eq:dvardef}) is simply related to the 3D Fourier power spectrum of $\rho$ by
\begin{equation} \label{eq:dvarspectrelation}
\dvar(\Sigma,\ell) \propto \pthreed(\rho,k) \propto \Sigma^2 \propto \ell^{-\alpha}\,
\end{equation}
for $\pthreed(\rho,k)\propto k^\alpha$. Thus, the $\Delta$-variance slope of $\Sigma$ is simply the negative slope of the power spectrum of $\rho$, if both follow power-law scalings. We will make use of this result in Section~\ref{sec:obs} below.

\begin{figure*}[t]
\centerline{
\includegraphics[width=0.95\linewidth]{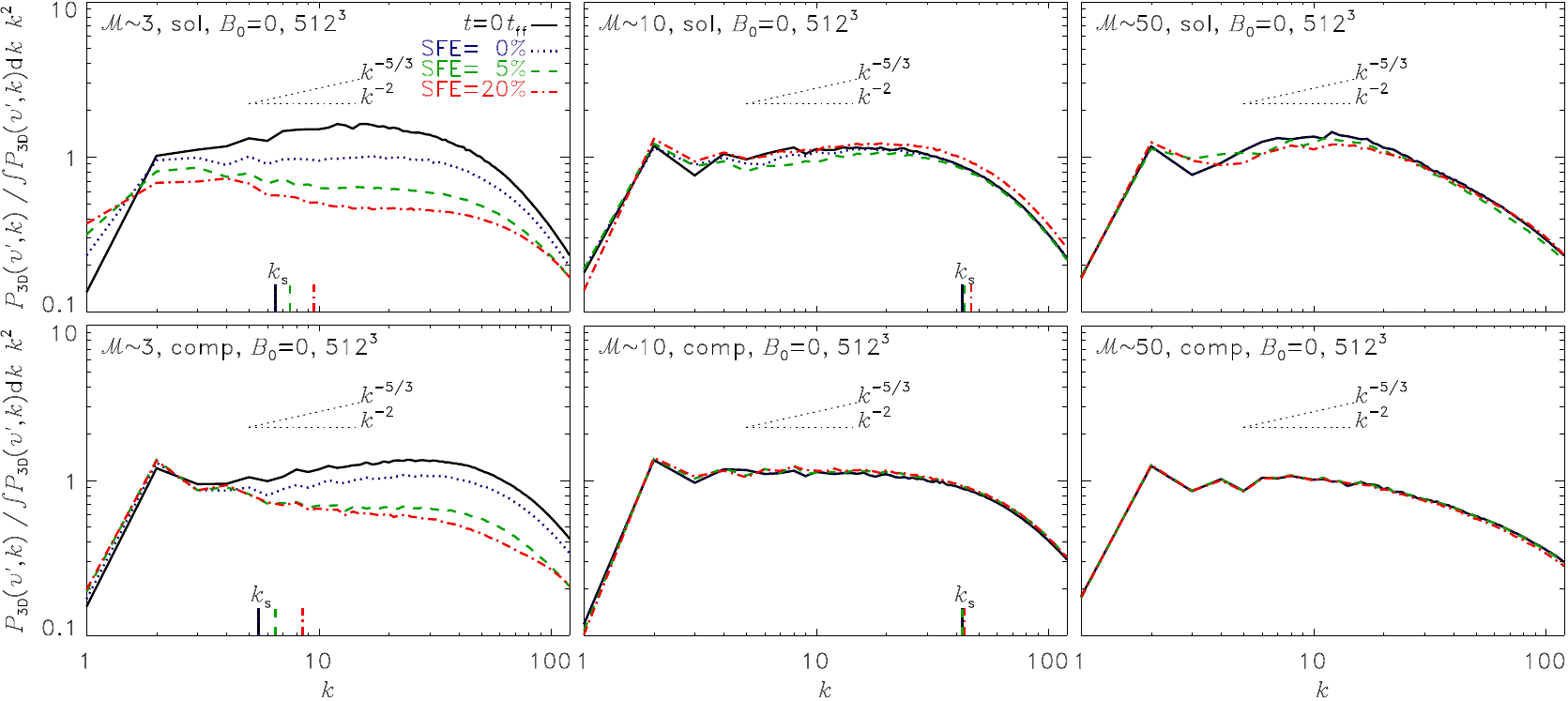}
}
\caption{Normalized velocity spectra of $q=\vect{v}/\cs$ defined in Equation~(\ref{eq:pthreed}) and compensated by $k^2$ for hydrodynamic simulations with resolution of $512^3$ grid cells and sonic RMS Mach number $\mach\sim3$ (\emph{left}), $\mach\sim10$ (\emph{middle}), and $\mach\sim50$ (\emph{right}) for solenoidal forcing (\emph{top}) and compressive forcing (\emph{bottom}). Different line styles indicate different times or SFEs: $t=0$ (the time when turbulence is fully developed and self-gravity is switched on; solid), $\sfe=0$ (right before the first sink particle forms; dotted), and $\sfe=5\%$ (dashed) and 20\% (dot-dashed). Straight dotted lines indicate Kolmogorov ($\propto k^{-5/3}$) or Burgers ($\propto k^{-2}$) scaling of turbulence in the approximate inertial range $5\lesssim k \lesssim 15$. Estimates of the sonic wave number $k_\mathrm{s}$ are shown as vertical lines on top of the abscissa. The sonic scale separates the range of supersonic turbulence ($k<k_\mathrm{s}$) from subsonic turbulence ($k>k_\mathrm{s}$).}
\label{fig:velspectra}
\end{figure*}

\subsection{Velocity Spectra} \label{sec:velspect}
We start by considering velocity spectra in order to compare them with previous results for supersonic turbulence, and to study their Mach-number and forcing dependences. Figure~\ref{fig:velspectra} shows power spectra of the velocity vector $\vect{v}$ normalized by the sound speed $\cs$, i.e., the local Mach number. Thus, we set $q=\vect{v}/\cs$ in Equation~(\ref{eq:pthreed}) and define the shortcut $v^\prime\equiv v/\cs$ . Since the simulations cover a wide range of RMS Mach numbers \mbox{$\mach\sim3$--$50$}, and hence the absolute power of each spectrum varies depending on $\mach$, we chose to normalize all velocity power spectra by their individual integral in Figure~\ref{fig:velspectra}, such that the shape of the spectra (and in particular their slopes) can be more easily compared between the different numerical models. Moreover, the spectra are compensated by $k^2$, such that a horizontal line would correspond to a power-law spectrum $\propto k^{-2}$ as in \citet{Burgers1948} turbulence. For comparison, we plot two dotted lines with power-law scaling $\propto k^{-5/3}$ \citep{Kolmogorov1941c} and $\propto k^{-2}$ \citep{Burgers1948}.

Before starting to analyze the dependence of the power spectra on the forcing, sonic Mach number and magnetic field, we first have to identify the scales on which the spectra are robust, numerically converged, and can be directly compared. At small wavenumbers, \mbox{$k=1$--$3$}, the turbulent forcing directly affects the spectra, because kinetic energy is injected there (see Section~\ref{sec:forcing}), which then self-consistently cascades down to smaller scales (larger wavenumbers). At the high-wavenumber end, numerical resolution and the bottleneck effect \citep{Falkovich1994,DoblerEtAl2003} affect the spectra. As shown in previous studies, scales smaller than about 30 grid cells, i.e., $k>k_\mathrm{diff}$ with $k_\mathrm{diff}\sim512/30\sim17$ for the $512^3$-resolution spectra analyzed here, are typically affected by numerical diffusion \citep{KitsionasEtAl2009,FederrathDuvalKlessenSchmidtMacLow2010,PriceFederrath2010,SurEtAl2010,FederrathSurSchleicherBanerjeeKlessen2011,KritsukEtAl2011Codes}. Moreover, the bottleneck effect could in principle affect the spectra at slightly smaller wavenumbers than $k_\mathrm{diff}$. However, previous studies indicate that the bottleneck effect has a minor impact on velocity spectra of supersonic turbulence \citep[unlike in mildly compressible or incompressible turbulence; see][]{DoblerEtAl2003}. Compressive forcing, in particular, does not seem to produce a significant bottleneck effect at all \citep[see the $1024^3$ resolution spectra of $\mach\sim5.5$ turbulence with solenoidal and compressive forcing in][]{FederrathDuvalKlessenSchmidtMacLow2010}.
This is likely because steeper spectra (e.g., $\propto k^{-2}$ as in highly compressible, supersonic turbulence) have a smaller correction due to the bottleneck effect than shallower spectra (e.g., $\propto k^{-5/3}$ as in incompressible turbulence) \citep{Falkovich1994}, and so the bottleneck effect is weaker in supersonic than in incompressible turbulence.
Thus, both forcing ($k<3$) and primarily numerical diffusion ($k\gtrsim17$) limit the scaling range in which all power spectra can be reasonably analyzed and compared. The remaining, intermediate range is quite small, but still allows us to draw general conclusions. In order to leave some tolerance on both ends, we define the approximate inertial range here as $5\lesssim k \lesssim 15$.

In Figure~\ref{fig:velspectra}, we see that the purely turbulent velocity spectra ($t=0$, solid lines) are roughly consistent with power laws, $\pthreed\propto k^{-5/3}$, for \citet{Kolmogorov1941c} turbulence below the sonic scale ($k>k_\mathrm{s}$), and are generally consistent with $\pthreed\propto k^{-2}$ on scales larger than the sonic scale ($k<k_\mathrm{s}$). We calculate the sonic scale as in \citet{SchmidtEtAl2009} and \citet{FederrathDuvalKlessenSchmidtMacLow2010} by evaluating the implicit integral definition $\int_{k_\mathrm{s}}^\infty \pthreed(v/\cs,k)\deriv k=1$. This definition states that the kinetic energy $\langle v^2\rangle/2$ on all scales $k>k_\mathrm{s}$ must be equal to the sound energy $\cs^2/2$ (note that here we did not include a factor of $1/2$ in the definition of the power spectrum, Equation~\ref{eq:pthreed}, so it simply cancels on both sides of the integral definition of $k_\mathrm{s}$). The sonic scale thus separates supersonic scales ($k<k_\mathrm{s}$) from subsonic scales ($k>k_\mathrm{s}$). We indicate $k_\mathrm{s}$ by the vertical lines on top of the abscissas in Figure~\ref{fig:velspectra}. It shifts to smaller scales (higher $k$) for models with increasing $\mach$, which simply means that a larger fraction of resolved wavenumbers remains in the supersonic regime. Hence, in the $\mach\sim50$ runs, the turbulence is supersonic on all scales, and the sonic wavenumber is larger than the cutoff wavenumber given by the numerical resolution. On the other hand, in the $\mach\sim3$ runs, the sonic scale is in the resolved range of scales, where the spectra are consistent with $\propto k^{-5/3}$ scaling for $k>k_\mathrm{s}$. Even though the spectra are quite noisy (because we only consider one realization of the turbulence at a fixed time) and the range of scales for analysis is limited as explained above, we find a flattening of the spectra from $P\propto k^{-2}$ in the supersonic regime ($k<k_\mathrm{s}$) to $P\propto k^{-5/3}$ in the subsonic regime ($k>k_\mathrm{s}$).

Compressive forcing exhibits slightly steeper spectra than solenoidal forcing for any given Mach number. This is because compressive forcing produces stronger shocks, and hence we expect a scaling closer to \citet{Burgers1948} scaling, $P\propto k^{-2}$. A slight steepening of velocity spectra was previously seen only in simulations with \mbox{$\mach\sim5$--$6$}, in which solenoidal forcing yielded a slope of $-1.86\pm0.05$ and compressive forcing a slope of $-1.94\pm0.05$ \citep{FederrathKlessenSchmidt2009,FederrathDuvalKlessenSchmidtMacLow2010}. The reason for the slightly steeper velocity spectra is that compressive forcing excites a larger fraction of compressible modes at any given $\mach$ \citep{FederrathEtAl2011PRL}, which---in the limit of a hypersonically turbulent velocity field, completely filled with shocks---is the essence of \cite{Burgers1948} scaling of turbulence.

Observations of molecular clouds support a velocity scaling between $k^{-5/3}$ (Kolmogorov) and $k^{-2}$ (Burgers), corresponding to $v\propto\ell^{1/3}$ and $v\propto\ell^{1/2}$ respectively, with most clouds exhibiting a scaling closer to Burgers turbulence \citep{Larson1981,SolomonEtAl1987,FalgaronePugetPerault1992,OssenkopfMacLow2002,HeyerBrunt2004,RomanDuvalEtAl2011}. Hence, both solenoidal and compressive forcing produce velocity scalings consistent with observations.

Our velocity spectra do not strongly change when star formation proceeds, as can be seen by comparing the $t=0$ and $\sfe=0$, 5\%, and 20\% curves in Figure~\ref{fig:velspectra}. Only the runs with $\mach\sim3$ show a significant dependence on $\sfe$, because in those models, global collapse starts to increase the velocity power on large scales, $k\sim1$, and thus reduces the relative power on smaller scales, because of the normalization of the spectra. Thus, velocity spectra do not seem to be a suitable indicator for star formation, unless the collapse is global, which is typically not observed \citep{ZuckermanPalmer1974,ZuckermanEvans1974,KrumholzTan2007,EvansEtAl2009}\footnote{See, however, the discussion in \citet{VazquezSemadeniEtAl2010} suggesting that global collapse drives turbulence and star formation. For instance, \citet{KlessenHennebelle2010} have argued and \citet{FederrathSurSchleicherBanerjeeKlessen2011} have shown that collapse and accretion do drive turbulence, but at the same time, the turbulence produced in this conversion from gravitational energy cannot hold the collapse. So global collapse would still be observed, unless the stars form much more quickly than global contraction occurs, adding an additional source of turbulence, possibly capable of holding the global collapse or even dispersing the cloud.}.

Our results provide an explanation for the insensitivity of the Principal Component Analysis of velocity fluctuations measured with CO lines in the comparison of Rosette (a star-forming cloud) and G216-2.5 (a quiescent cloud) found in \citet{HeyerWilliamsBrunt2006}. CO traces primarily the large-scale structure of molecular clouds, because the \mbox{1--0} and \mbox{2--1} rotational transition lines become optically thick at relatively low column density for both $^{12}\mathrm{CO}$ and $^{13}\mathrm{CO}$. The results in Figure~\ref{fig:velspectra} indeed indicate that large-scale velocity analyses of a molecular cloud are not particularly sensitive to star formation. In the next section, however, we show that the density and column density power spectra do depend on star formation, even on relatively large scales, and correlate with the $\sfe$.

\begin{figure*}[t]
\centerline{
\includegraphics[width=0.95\linewidth]{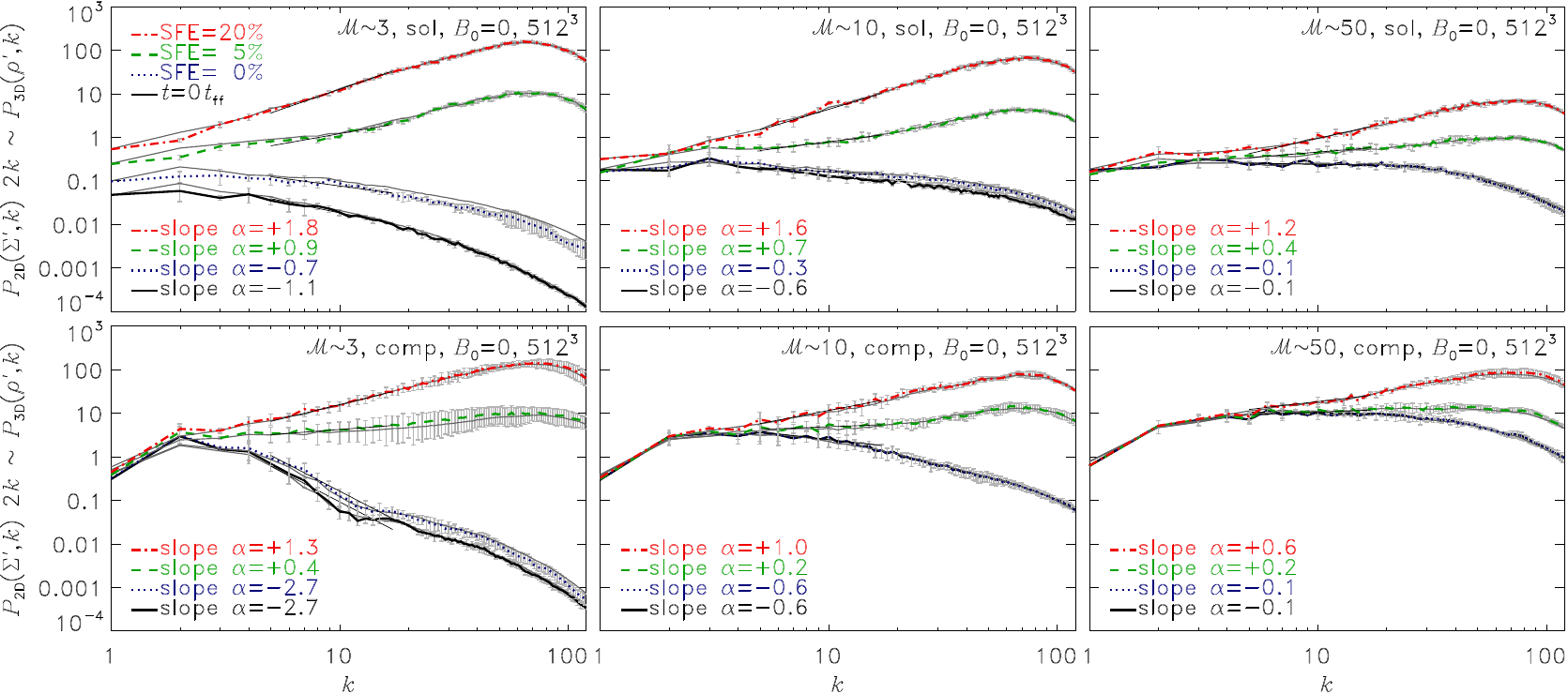}
}
\caption{Same as Figure~\ref{fig:velspectra}, but for the column density fluctuation spectra $\ptwod(\Sigma^\prime,k)$ with $q=\Sigma/\Sigma_0-1$ in Equation~(\ref{eq:ptwod}) and multiplied by $2k$ (thick lines) to transform the column density spectra to volumetric density spectra $\pthreed(\rho^\prime,k)$ with $q=\rho/\meanrho-1$ in Equation~(\ref{eq:pthreed}) (thin lines). Both 2D and 3D spectra agree as expected from Equation~(\ref{eq:ptwodtothreedrelation}), within the uncertainties introduced by different projection directions along the $x$, $y$, or $z$ axes, indicated by error bars. Power-law fits in the approximate inertial range, $5\leq k\leq 15$, show that the density spectra evolve from negative power-law slope $\alpha$ defined in Equation~(\ref{eq:alpha}) to a positive slope, when star formation proceeds in the clouds ($\sfe\gtrsim0$), irrespective of the particular model parameters. This makes the density and column density spectra a powerful tool to distinguish the evolutionary state of a cloud and to estimate the SFE in observations, as will be discussed in Sections~\ref{sec:estimatingsfe} and~\ref{sec:obs} below.}
\label{fig:coldensspectra}
\end{figure*}

\begin{figure*}[t]
\centerline{
\includegraphics[width=0.95\linewidth]{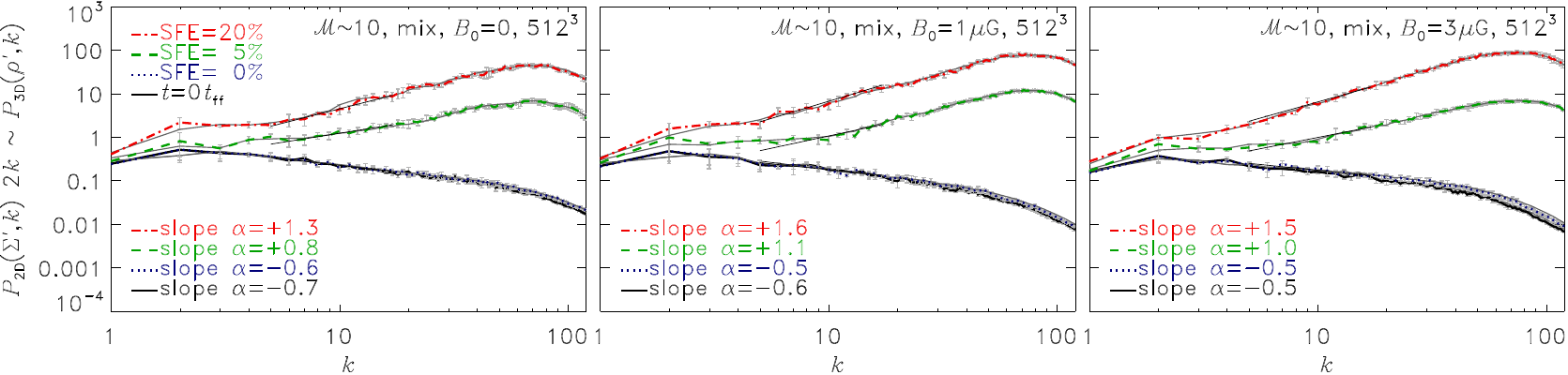}
}
\caption{Same as Figure~\ref{fig:coldensspectra}, but for $\mach\sim10$ models with mixed forcing and varying magnetic field strength $B_0=0$ (\emph{left}), $1\,\mu\Gauss$ (\emph{middle}), and $3\,\mu\Gauss$ (\emph{right}). The slope $\alpha$ of the density spectra changes from negative to positive for star-forming clouds ($\sfe\gtrsim0$), irrespective of the magnetic field.}
\label{fig:Bspectra}
\end{figure*}

\subsection{Density and Column Density Spectra} \label{sec:rhospect}

For the density and column density spectra, we define $\rho^\prime\equiv\rho/\meanrho-1$ and $\Sigma^\prime\equiv\Sigma/\Sigma_0-1$ with the mean density $\meanrho$ and the mean column density $\Sigma_0$. We then set $q=\rho^\prime$ in Equation~(\ref{eq:pthreed}), which defines $\pthreed(\rho^\prime,k)$, and set $q=\Sigma^\prime$ in Equation~(\ref{eq:ptwod}), defining $\ptwod(\Sigma^\prime,k)$. The normalization by the mean values is useful, because it enables a direct comparison of the spectra, even when the absolute density and column density scales are different in the various simulation models (see Table~\ref{tab:sims}).

Figure~\ref{fig:coldensspectra} shows the same as Figure~\ref{fig:velspectra}, but for the density and column density fluctuation spectra. Note that we expect $\pthreed(\rho^\prime,k)\sim2k\ptwod(\Sigma^\prime,k)$ from Equation~(\ref{eq:ptwodtothreedrelation}), which is well confirmed in Figure~\ref{fig:coldensspectra}, showing $2k\ptwod(\Sigma^\prime,k)$ as thick lines and $\pthreed(\rho^\prime,k)$ as thin lines superimposed. The match is quite close, as expected for nearly isotropic, turbulent density distributions. The gray error bars indicate variations of $\ptwod(\Sigma^\prime,k)$ with different projection directions along the $x$, $y$, and $z$ axes, showing that there is no significant variation of the column density spectra with projection direction.

We plot density spectra for different times and SFEs, $t=0$ and $\sfe=0$, $5\%$, and $20\%$ in Figure~\ref{fig:coldensspectra}. We see that the dependence of $\pthreed(\rho^\prime,k)$ and $\ptwod(\Sigma^\prime,k)$ on $\sfe$ is very strong. All spectra have a negative slope $\alpha$ in the purely turbulent regime before self-gravity is considered in the simulations ($t=0$) until the first sink particle forms ($\sfe=0$). We here quote the slope $\alpha$ of the 3D density spectrum by fitting a power law within the approximate inertial range, $5\leq k\leq 15$, with $\pthreed(\rho^\prime,k)\propto k^\alpha$. Note that the column density power spectrum $\ptwod(\Sigma^\prime,k) \propto k^{\alpha-1}$ (see Equation~\ref{eq:ptwodtothreedrelation}), which is why we plot $2k\ptwod$, also exhibiting the slope $\alpha$. For $\sfe=5\%$, this slope becomes positive and increases further for $\sfe=20\%$ in all numerical models, irrespective of the forcing and sonic Mach number. This result and our measurements of the slope $\alpha$ do not depend on numerical resolution or statistical sampling of the turbulence as demonstrated in Figures~\ref{fig:coldensspectra_res} and~\ref{fig:pdfspectseeds} in the Appendix.

The difference of $\alpha$ between $t=0$ and $\sfe=20\%$ is largest in the $\mach\sim3$ runs, intermediate for $\mach\sim10$, and smallest in the $\mach\sim50$ simulations. Comparing different forcing, we see that on average, solenoidal forcing produces systematically larger slopes $\alpha$ than compressive forcing at any given Mach number. This has been noticed previously by \citet{FederrathKlessenSchmidt2009}, but only for \mbox{$\mach\sim5$--6} simulations in the purely turbulent regime, and is related to the different fractal dimensions $\Df $ obtained for solenoidal forcing ($\Df\sim2.6$) and compressive forcing ($\Df\sim2.3$). Here we see that the higher the Mach number, the smaller are the differences between solenoidal and compressive forcing, which is plausible, because we expect both forcings to produce a statistically identical network of pure shocks for $\mach\to\infty$, which is the theoretical limit of Burgers turbulence. We also see that in this limit, the differences in slope with time and $\sfe$ become smaller, because gas at very high Mach numbers is so strongly compressed in shocks and dense filaments, that collapse does not change the density structure as much as in the case of lower Mach numbers. However, the fact that the slope $\alpha$ is negative for $\sfe\sim0$ and becomes positive for $\sfe\gtrsim0$ is robust, even when the Mach number is increased from $\mach\sim3$ to 50, and even when the forcing is varied from one extreme (solenoidal forcing) to the opposite extreme (compressive forcing).

In Figure~\ref{fig:Bspectra}, we see the same general trend of increasing slope $\alpha$ with $\sfe$ in the MHD simulations with $\mach\sim10$ and mixed forcing. The density power spectra do not seem to depend systematically on the magnetic field strength, consistent with the conclusion drawn in \citet{CollinsEtAl2012}. The spectra for $B=1\,\mu\Gauss$ and $3\,\mu\Gauss$ have a slope $\alpha$ that is slightly larger (by about 0.2 on average) than in the $B=0$ case, which is a relatively small, but noticeable difference. However, the change in slope from $\alpha<0$ for $\sfe\sim0$ to $\alpha>0$ for $\sfe\gtrsim0$ is clearly seen in the MHD models in Figure~\ref{fig:Bspectra}, as before in Figure~\ref{fig:coldensspectra} for different forcing and sonic Mach number.

All density spectra in Figures~\ref{fig:coldensspectra} and~\ref{fig:Bspectra} are broadly consistent with previous studies exploring a limited subset of the parameter space analyzed in this study \citep{BeresnyakEtAl2005,KowalLazarianBeresnyak2007,KritsukEtAl2007,FederrathKlessenSchmidt2009,CollinsEtAl2012}. Here we show that all density spectra exhibit a flattening with increasing Mach number for both solenoidal and compressive forcing, consistent with a previous study of purely solenoidal forcing \citep{KimRyu2005}. However, the most striking result is that all density spectra exhibit negative power-law slopes for $\sfe\sim0$, but turn to positive slopes for $\sfe>0$, making them a potentially powerful tool to distinguish star-forming clouds ($\sfe>0$) from clouds that did not (yet) form stars ($\sfe=0$). Moreover, the slope of the density spectra increases roughly monotonically with increasing $\sfe$ (explored in more detail below). Here we show that the switching in slope from negative to positive $\alpha$ of the density and column density spectra is universal, holding in clouds with \mbox{$\mach=3$--$50$} for solenoidal and compressive forcing, as well as for different magnetic field strengths. This universal behavior of the density spectrum is a signpost of gravitational collapse when star formation proceeds. As more and more power in the density field moves to small scales where star formation occurs, the density spectrum rises on those scales and causes the significant change of the slope $\alpha$ that we see here.

\begin{figure*}[t]
\centerline{
\includegraphics[width=0.95\linewidth]{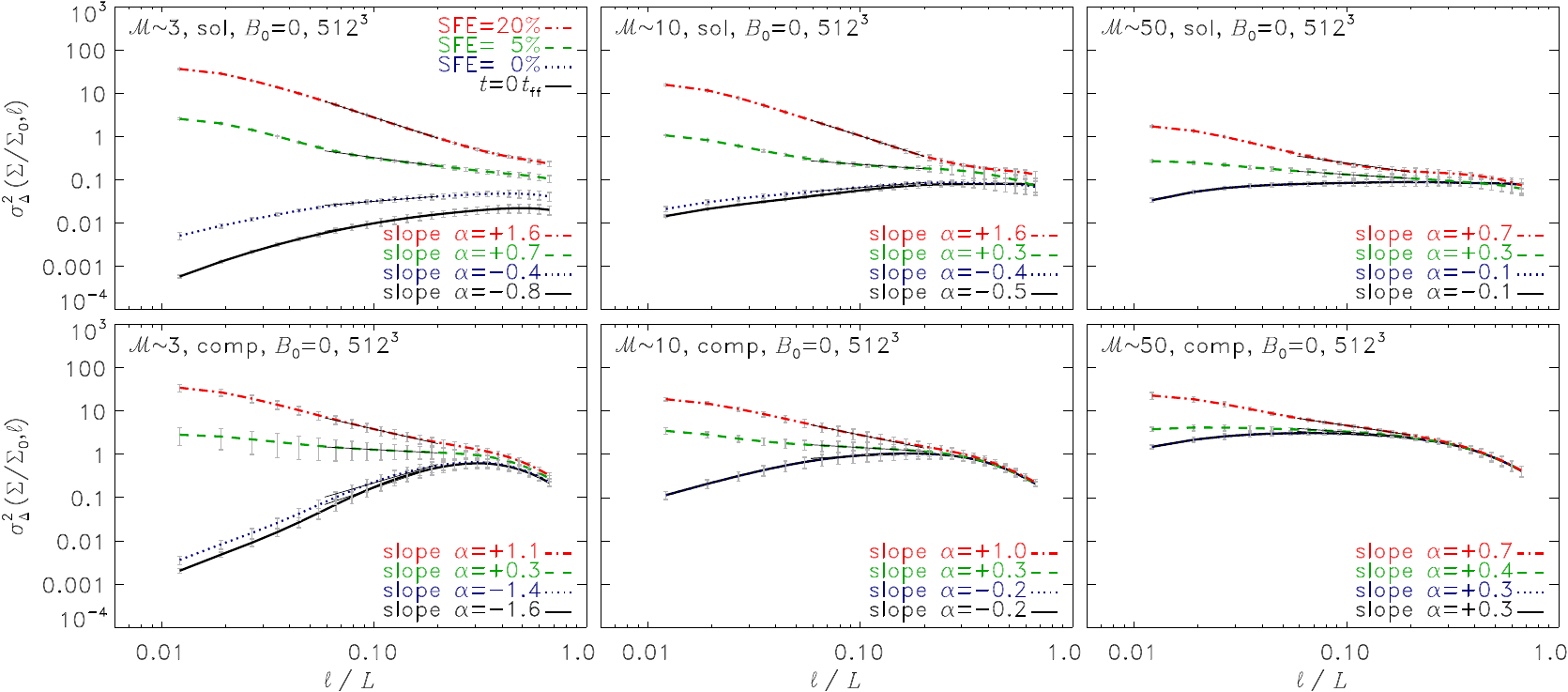}
}
\caption{Same as Figure~\ref{fig:coldensspectra}, but for the $\Delta$-variance spectra of the column density contrast, $\dvar(\Sigma/\Sigma_0,\ell)$, defined in Equation~(\ref{eq:dvardef}). Note that the slopes $\alpha$ quoted in the legend were measured in the approximate inertial range, $0.06\leq\ell/L\leq 0.2$, and recast into the slope of the 3D density Fourier spectrum, $\pthreed(\rho,k)\propto k^\alpha$. The slopes computed from the $\Delta$-variance spectra are similar to the slopes in Figure~\ref{fig:coldensspectra}, as expected for approximate power laws (see Equation~\ref{eq:dvarspectrelation}).}
\label{fig:deltavar}
\end{figure*}

\begin{figure*}[t]
\centerline{
\includegraphics[width=0.95\linewidth]{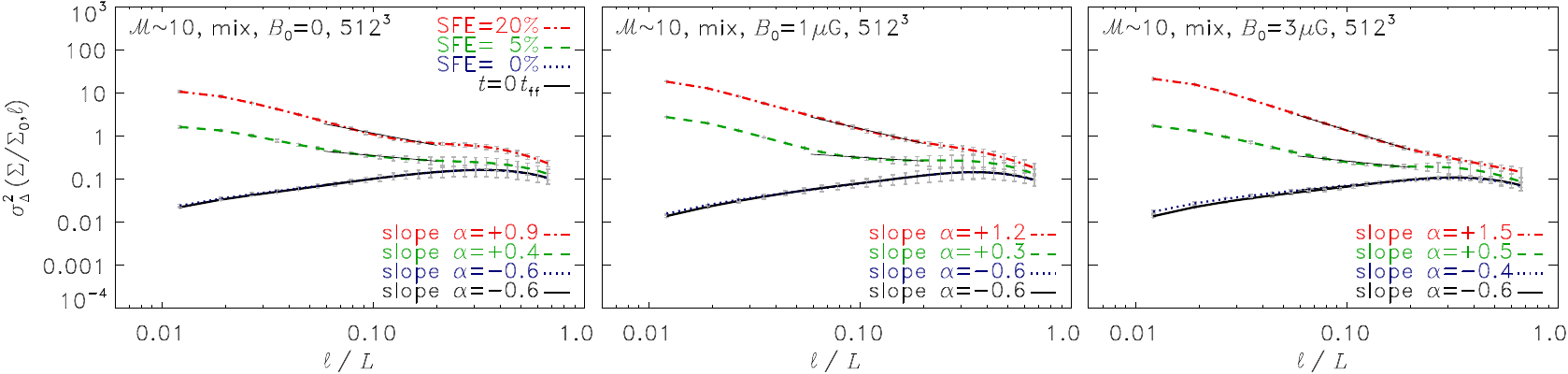}
}
\caption{Same as Figure~\ref{fig:deltavar}, but for $\mach\sim10$ models with mixed forcing and varying magnetic field strength $B_0=0$ (\emph{left}), $1\,\mu\Gauss$ (\emph{middle}), and $3\,\mu\Gauss$ (\emph{right}). As for the hydrodynamic models of Figure~\ref{fig:deltavar}, the MHD models indicate a strong dependence of the slope $\alpha$ on $\sfe$.}
\label{fig:Bdeltavar}
\end{figure*}

\subsection{Delta-variance Spectra of the Column Density} \label{sec:dvar}

As a more practical approach to measure the density power spectrum in molecular cloud observations, we now study the $\Delta$-variance spectra, defined in Section~\ref{sec:dvardef}. The $\Delta$-variance tool provided by \citet{OssenkopfKripsStutzki2008a} can treat spatial variations in the signal-to-noise ratio, as well as non-periodic boundaries, making it a useful tool to measure the column density spectrum in observational maps. For pure power-law distributions, the $\Delta$-variance of the column density, $\dvar(\Sigma,\ell)$ gives information equivalent to Fourier spectra \citep{StutzkiEtAl1998}, and is related to the 3D density power spectrum $\pthreed(\rho,k)$ via Equation~(\ref{eq:dvarspectrelation}).

Figure~\ref{fig:deltavar} shows the same models and times as in Figure~\ref{fig:coldensspectra}, but for the $\Delta$-variance spectra of the column density contrast $\dvar(\Sigma/\Sigma_0,\ell)$, where we divide by the mean column density $\Sigma_0$ to facilitate the comparison between different models. As for the Fourier spectra of the density and column density in Figure~\ref{fig:coldensspectra}, we see that the slope of $\dvar(\Sigma/\Sigma_0,\ell)$ depends sensitively on local collapse and star formation. A similar trend was reported in simulations analyzed by \citet{OssenkopfKlessenHeitsch2001}, yet with a much more limited dynamic range and set of parameters. The basic effect was already seen in that study, but \citet{OssenkopfKlessenHeitsch2001} did not relate the slope $\alpha$ to the $\sfe$, because of the relatively large uncertainties in their simulations. For the present set of simulations, however, we can determine the slopes $\alpha$ reliably in the approximate inertial range, $0.06\leq\ell/L\leq 0.2$, without being affected by numerical resolution issues (see Figure~\ref{fig:coldensspectra_res} in the Appendix).

We recast the slope of the $\Delta$-variance into the slope $\alpha$ of $\pthreed(\rho,k)$ via Equation~(\ref{eq:dvarspectrelation}) to make them directly comparable to the Fourier spectrum slopes. Comparing the slopes $\alpha$ in Figures~\ref{fig:coldensspectra} and~\ref{fig:deltavar}, we find good agreement. The difference in slopes between the $\Delta$-variance and the Fourier spectra is typically less than $\pm0.3$, caused by statistical uncertainties in the fit range and by the fact that the inertial range is only approximately described by a power law. However, the changes in slope with increasing $\sfe$ are clearly significant and consistent with the results obtained in Figure~\ref{fig:coldensspectra}.

Moreover, we find that the $\Delta$-variance slopes of runs with different magnetic field also agree well with the Fourier spectra shown in Figure~\ref{fig:Bspectra}. The general conclusion from all spectra is thus robust: the slope $\alpha$ switches sign from negative to positive when star formation proceeds ($\sfe\gtrsim0$), irrespective of the sonic Mach number, the forcing, or the magnetic field strength.

\section{Estimating the SFE} \label{sec:estimatingsfe}

\begin{figure}[t]
\centerline{
\includegraphics[width=0.99\linewidth]{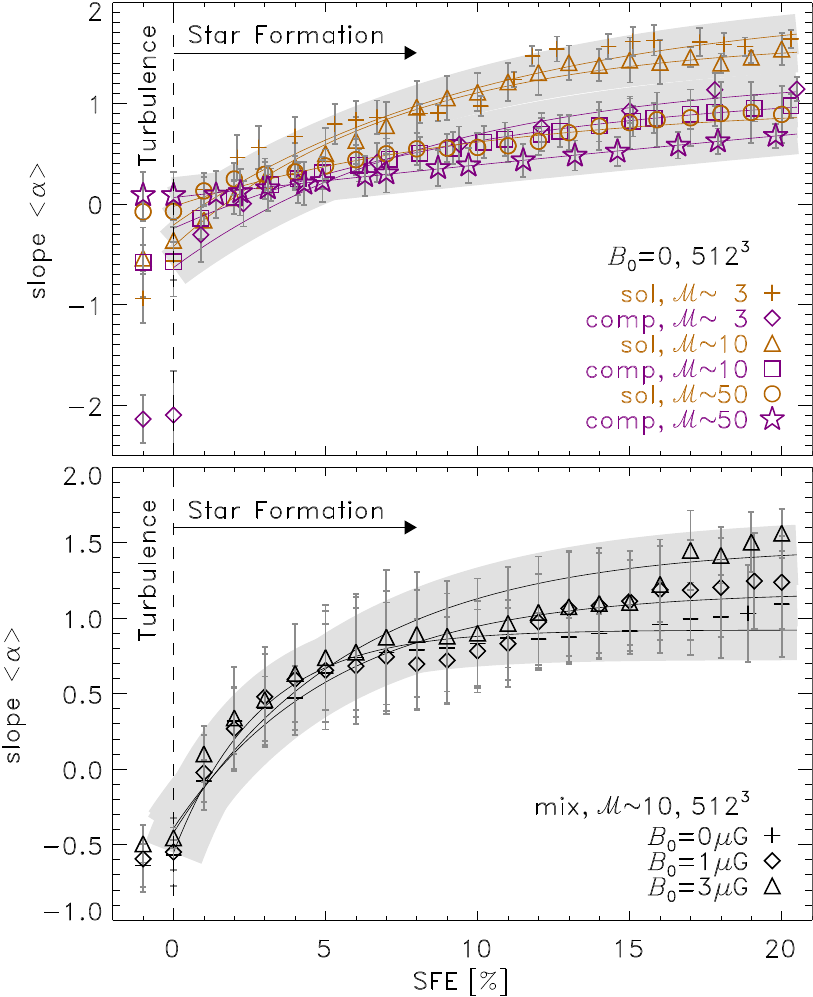}
}
\caption{Spectral slope $\langle\alpha\rangle$, averaged between the Fourier spectra (Figures~\ref{fig:coldensspectra} and~\ref{fig:Bspectra}) and the $\Delta$-variance analysis (Figures~\ref{fig:deltavar} and~\ref{fig:Bdeltavar}) as a function of $\sfe$. The \emph{top panel} shows hydrodynamical models with solenoidal and compressive forcing at sonic $\mach\sim3$, 10, and 50, while the \emph{bottom panel} shows MHD models of mixed forcing at $\mach\sim10$, but with different magnetic field strength $B_0=0$, $1\,\mu\Gauss$, and $3\,\mu\Gauss$. Error bars include the uncertainties from three different projections (along $x$, $y$, and $z$) and the uncertainties from averaging between Fourier and $\Delta$-variance spectra. The vertical dashed line at $\sfe=0$ separates the turbulent regime from the regime of star formation ($\sfe>0$). Overall, the slope $\alpha$ monotonically increases with $\sfe$ in each model. Model fits using Equation~(\ref{eq:fit}) with fit parameters listed in Table~\ref{tab:fits} are shown as solid lines. Gray shaded regions indicate deviations from the fits by $\alpha\pm0.2$.}
\label{fig:slopes}
\end{figure}

In the previous section, we showed that the spectral slope $\alpha$ of all density power spectra switches from negative to positive when star formation proceeds. We now study the quantitative dependence of $\alpha$ on $\sfe$. Figure~\ref{fig:slopes} shows $\langle\alpha\rangle$ averaged between the Fourier spectra (Figures~\ref{fig:coldensspectra} and~\ref{fig:Bspectra}) and the $\Delta$-variance spectra (Figures~\ref{fig:deltavar} and~\ref{fig:Bdeltavar}) as a function of $\sfe$. We see that $\alpha$ increases with $\sfe$ from $\alpha<0$ for $\sfe\sim0$ to $\alpha>0$ for $\sfe>0$, except for the $\mach\sim50$ models (top panel), where $\alpha\sim0$ for $\sfe\sim0$. These high-Mach number models also show the slowest increase, from $\alpha\sim0$ for $\sfe\sim0$ to $\alpha\sim0.8$ for $\sfe=20\%$. The $\mach\sim3$ models, on the other hand, show a much stronger difference between $\sfe\sim0$ and $\sfe>0$. This can be explained by the fact that the density structure changes significantly during collapse in the low-Mach number models, while the global structure is almost unaffected during star formation in the high-Mach number cases, because the latter already contain most of the mass in high-density filaments caused by turbulent compression (see Figures~\ref{fig:imagessol} and~\ref{fig:imagescomp}).

For a given sonic Mach number, we see that solenoidal forcing always produces higher $\alpha$ than compressive forcing (top panel in Figure~\ref{fig:slopes}). However, the differences between the forcings become smaller as the Mach number increases. This is quite plausible, because in the hypersonic limit ($\mach\to\infty$), we expect both solenoidal and compressive forcing to produce maximum compression, resulting in a network of pure shocks as in \citet{Burgers1948} turbulence. The differences between solenoidal and compressive forcing with $\mach\sim50$ are indeed not significant, indicated by the overlapping error bars in those two cases (circles and stars). Real molecular clouds have sonic Mach numbers in the range \mbox{$\mach\sim2$--$50$}, with most clouds exhibiting $\mach\sim10$, where differences between the forcings become pronounced when $\sfe\gtrsim5\%$ (triangles and squares). For Mach numbers $\mach\sim3$ (crosses and diamonds), the differences between solenoidal and compressive forcing are significant in both the purely turbulent ($\sfe\sim0$) and star-forming regimes ($\sfe>0$).

The bottom panel of Figure~\ref{fig:slopes} shows that adding a magnetic field does not change $\alpha$ significantly, except in the late stages of star formation when $\sfe>10\%$. There, the higher the magnetic field, the slightly larger becomes $\alpha$.

\begin{table}
\caption{Fit Parameters for Each Model in Figure~\ref{fig:slopes}.}
\label{tab:fits}
\def\arraystretch{1.2}
\setlength{\tabcolsep}{8pt}
\begin{tabular}{r|ccc}
\hline
\hline
\multicolumn{1}{l}{HD models (Figure~\ref{fig:slopes} top)} & $p_0$ & $p_1$ & $p_2$ \\
\hline
solenoidal, $\mach\sim\phantom{0}3$, $B_0=0\,\mu\Gauss$  & 1.990  & 0.788  & 0.097 \\
compressive, $\mach\sim\phantom{0}3$, $B_0=0\,\mu\Gauss$  & 1.288  & 0.653  & 0.117 \\
solenoidal, $\mach\sim10$, $B_0=0\,\mu\Gauss$  & 1.633  & 0.721  & 0.136 \\
compressive, $\mach\sim10$, $B_0=0\,\mu\Gauss$  & 1.149  & 0.332  & 0.097 \\
solenoidal, $\mach\sim50$, $B_0=0\,\mu\Gauss$  & 0.988  & 0.014  & 0.099 \\
compressive, $\mach\sim50$, $B_0=0\,\mu\Gauss$  & 3.437  & 1.216  & 0.010 \\
\hline
\multicolumn{1}{l}{MHD models (Figure~\ref{fig:slopes} bottom)} & $p_0$ & $p_1$ & $p_2$ \\
\hline
mixed, $\mach\sim10$, $B_0=0\,\mu\Gauss$ & 0.921  & 0.389  & 0.356 \\
mixed, $\mach\sim10$, $B_0=1\,\mu\Gauss$ & 1.175  & 0.450  & 0.192 \\
mixed, $\mach\sim10$, $B_0=3\,\mu\Gauss$ & 1.478  & 0.642  & 0.169 \\
\hline
\end{tabular}

\textbf{Notes.} The model function for fitting is given by Equation~(\ref{eq:fit}).
\end{table}

In order to better distinguish models with different physical parameters and to provide a quantitative measure of the relation between $\alpha$ and $\sfe$, we made empirical fits to the data in Figure~\ref{fig:slopes}. We use the following model function for fitting,
\begin{equation} \label{eq:fit}
\alpha=p_0-\exp\left(p_1-p_2\times\sfe[\%]\right)\,,
\end{equation}
with parameters $p_0$, $p_1$, and $p_2$, because of its simplicity and because it can be inverted. The positive-definite inverse function
\begin{equation} \label{eq:fitinverse}
\sfe[\%] = \mathrm{max}\left\{0,\,\left[p_1-\ln\left(p_0-\alpha\pm0.2\right)\right] / p_2\right\}
\end{equation}
thus provides an estimate of the $\sfe$ when $\alpha$ is measured from an observational map. We included an uncertainty in $\alpha\pm0.2$, reflecting the typical uncertainty of $\alpha$ in the simulations to establish the relation in Equation~(\ref{eq:fitinverse}). In practice, the uncertainties in $\alpha$ are likely somewhat higher than $\pm0.2$ when applied to real observations, so any further uncertainties (of systematic or statistical nature) must be added accordingly. Sources of such uncertainties and limitations of the present approach are discussed in Section~\ref{sec:uncertainties}. The fit parameters $p_0$, $p_1$, and $p_2$ are listed in Table~\ref{tab:fits} for each numerical model in Figure~\ref{fig:slopes}. The fit curves are shown in Figure~\ref{fig:slopes} as solid lines. Gray shaded regions centered on the best-fit curves indicate an uncertainty range of $\pm0.2$ in $\alpha$.

The dependence of $\alpha$ on $\sfe$ seen in Figure~\ref{fig:slopes} and provided by Equation~(\ref{eq:fit}) with the fit parameters in Table~\ref{tab:fits} suggests that by measuring the spectral slope $\alpha$ in an observational map of dust or molecular column density (e.g., by using the $\Delta$-variance technique described in Section~\ref{sec:dvardef}), one can distinguish star-forming ($\sfe>0$) from quiescent (purely turbulent) clouds ($\sfe\sim0$). Moreover, if an independent estimate of the sonic Mach number is available \citep[e.g.,][]{BurkhartEtAl2009} and if the forcing can be constrained \citep[][]{BruntFederrathPrice2010a,BruntFederrathPrice2010b,Brunt2010,KonstandinEtAl2012ApJ,KainulainenTan2012,Seon2012}, measuring $\alpha$ and comparing it with the present set of numerical simulations may give a hint on the SFE of an observed cloud, by inverting the curves shown in Figure~\ref{fig:slopes}. This inversion is given by Equation~(\ref{eq:fitinverse}) with the fit parameters in Table~\ref{tab:fits}. Similarly, if $\sfe$ and $\mach$ were known for a given cloud, Figure~\ref{fig:slopes} could be used to constrain the forcing in that cloud.

\section{Comparison with Observations} \label{sec:obs}

Figure~\ref{fig:slopes} showed that the spectral slope $\alpha$ is sensitive to the $\sfe$. Thus, by measuring that slope in interstellar cloud regions, we can constrain the $\sfe$ of observed clouds. Table~\ref{tab:obs} lists estimates of $\alpha$ measured with different tracers and for different cloud regions, taken from various sources in the literature. We find slopes in the range $\alpha\sim-1.6$ to +1.6, and sort each observation in Table~\ref{tab:obs} by $\alpha$ in ascending order (from top to bottom). For each of the following observational estimates, we include both the uncertainties given by the inversion, Equation~(\ref{eq:fitinverse}), combined with the fit uncertainties from the observation quoted in Table~\ref{tab:obs} (see also the discussion in Section~\ref{sec:uncertainties} on other uncertainties and limitations). Our observational estimates of the $\sfe$ are thus conservative and strictly yield only upper or lower limits for the $\sfe$.

\begin{table}
\caption{Estimates of the Slope $\alpha$ of the Density Power Spectrum $\pthreed(\rho,k)\propto k^\alpha$ and Inferred SFEs for Galactic Clouds.}
\label{tab:obs}
\def\arraystretch{1.2}
\setlength{\tabcolsep}{2.2pt}
\begin{tabular}{lcccr}
\hline
\hline
Cloud/Region & Ref. & Tracer & $\phantom{-}\alpha$ & SFE \\
\hline
01) Ursa Major cirrus & (1) & $\mathrm{HI}\;21\,\cm$ & $-1.6\pm0.2$ & $\sim\!0\%$ \\
02) Polaris Flare (IRAM) & (2) & $^{12}\mathrm{CO}\;1\!\to\!0$ & $-1.3\pm0.2$ & $\sim\!0\%$ \\
03) Polaris Flare (KOSMA) & (2) & $^{12}\mathrm{CO}\;2\!\to\!1$ & $-1.1\pm0.2$ & $\sim\!0\%$ \\
04) Polaris Flare (FCRAO) & (2) & $^{13}\mathrm{CO}\;1\!\to\!0$ & $-1.0\pm0.2$ & $\sim\!0\%$ \\
05) Polaris Flare (IRAM) & (2) & $^{12}\mathrm{CO}\;2\!\to\!1$ & $-0.9\pm0.2$ & $<\!1\%$ \\
06) Vela & (3) & dust extinc. & $-0.9\pm0.2$ & $<\!1\%$ \\
07) Polaris Flare & (4) & $^{13}\mathrm{CO}\;1\!\to\!0$ & $-0.8\pm0.1$ & $<\!1\%$ \\
08) Perseus/NGC 1333 & (2) & $^{13}\mathrm{CO}\;1\!\to\!0$ & $-0.8\pm0.2$ & $<\!1\%$ \\
09) Coalsack & (3) & dust extinc. & $-0.8\pm0.3$ & $<\!2\%$ \\
10) Orion A & (2) & $^{13}\mathrm{CO}\;1\!\to\!0$ & $-0.7\pm0.2$ & $<\!2\%$ \\
11) Orion B & (2) & $^{13}\mathrm{CO}\;1\!\to\!0$ & $-0.6\pm0.2$ & $<\!2\%$ \\
12) Polaris Flare (CfA) & (2) & $^{12}\mathrm{CO}\;1\!\to\!0$ & $-0.6\pm0.2$ & $<\!2\%$ \\
13) Lupus & (3) & dust extinc. & $-0.6\pm0.2$ & $<\!2\%$ \\
14) Mon OB1 & (3) & dust extinc. & $-0.6\pm0.2$ & $<\!2\%$ \\
15) Mon OB1/NGC 2264 & (3) & dust extinc. & $-0.5\pm0.1$ & $<\!2\%$ \\
16) Rosette & (3) & dust extinc. & $-0.5\pm0.3$ & $<\!4\%$ \\
17) IC 5146 & (3) & dust extinc. & $-0.4\pm0.2$ & $<\!4\%$ \\
18) Mon R2 & (2) & $^{13}\mathrm{CO}\;1\!\to\!0$ & $-0.4\pm0.3$ & $<\!4\%$ \\
19) Mon OB1/NGC 2264 & (2) & $^{13}\mathrm{CO}\;1\!\to\!0$ & $-0.4\pm0.3$ & $<\!4\%$ \\
20) Orion A+B & (3) & dust extinc. & $-0.3\pm0.2$ & $<\!4\%$ \\
21) Pipe & (3) & dust extinc. & $-0.3\pm0.2$ & $<\!4\%$ \\
22) Corona Australis & (3) & dust extinc. & $-0.3\pm0.3$ & $<\!5\%$ \\
23) Perseus & (3) & dust extinc. & $-0.2\pm0.2$ & $<\!5\%$ \\
24) Mon R2 & (3) & dust extinc. & $-0.1\pm0.2$ & $<\!6\%$ \\
25) Taurus & (3) & dust extinc. & $\phantom{-}0.0\pm0.2$ & $<\!10\%$\\
26) W3 & (3) & dust extinc. & $\phantom{-}0.0\pm0.2$ & $<\!10\%$ \\
27) Chamaeleon & (3) & dust extinc. & $\phantom{-}0.0\pm0.2$ & $<\!10\%$ \\
28) Cygnus X & (3) & $^{13}\mathrm{CO}\;1\!\to\!0$ & $\phantom{-}0.2\pm0.2$ & $<\!17\%$ \\
29) Serpens core & (5) & dust emiss. & $\phantom{-}1.6\pm0.4$ & $>\!8\%$ \\
\hline
\end{tabular}

\textbf{Notes.} The upper and lower limits of the SFE quoted in the last column were estimated based on comparing the range of spectral slopes $\alpha$ (including the maximum uncertainties in Equation~\ref{eq:fitinverse} combined with the additional fit uncertainties for $\alpha$ from the observations; see discussion in Section~\ref{sec:uncertainties}) with the numerical simulations in Figure~\ref{fig:slopes}, covering a wide range of physical conditions (\mbox{$\mach\sim3$--50} with solenoidal and compressive forcing, and different magnetic field strengths). Equation~(\ref{eq:fitinverse}) together with Table~\ref{tab:fits} provide direct relations $\sfe(\alpha)$ fitted to the simulation data in Figure~\ref{fig:slopes}. References: (1) Fourier analysis by \citet{MivilleDescheneEtAl2003}; (2) Reanalysis of $\Delta$-variance estimates in \citet{BenschStutzkiOssenkopf2001}; (3) Reanalysis of $\Delta$-variance estimates in \citet{SchneiderEtAl2011}; (4) $\Delta$-variance by \citet{StutzkiEtAl1998}; (5) $\Delta$-variance by \citet{OssenkopfKlessenHeitsch2001}.
\end{table}

The first entry in Table~\ref{tab:obs} is from a Fourier analysis of the $21\,\cm$ line of atomic hydrogen of the high Galactic latitude Ursa Major `cirrus' by \citet{MivilleDescheneEtAl2003}. They find a spectral slope $\alpha\sim-1.6\pm0.2$. Comparing this value, including the uncertainties in both simulations (Figure~\ref{fig:slopes} and Equation~\ref{eq:fitinverse}) and observations (Table~\ref{tab:obs}), we estimate that the Ursa Major cirrus has an $\sfe\sim0$. This is because even the maximum slope, $\alpha\sim-1.4$ is significantly negative and any of our numerical simulations did not produce stars for $\alpha\lesssim-1$. Although molecular condensations have been detected in the Ursa Major cirrus, it appears to have essentially no star formation activity, likely because of the high virial parameters estimated for this cloud \citep{DeVriesEtAl1987}. This is in good agreement with our findings of $\sfe\sim0$, based on the comparison of the observed slope $\alpha$ with our numerical simulations.

Entries 2--5, 7, and 12 in Table~\ref{tab:obs} are for observations of the Polaris Flare by \citet{StutzkiEtAl1998} and \citet{BenschStutzkiOssenkopf2001} with different instruments (IRAM, KOSMA, FCRAO, CfA) and different molecular transitions of $^{12}\mathrm{CO}$ and $^{13}\mathrm{CO}$. We see some variation of the derived slope $\alpha\sim-1.3$ to $-0.6$, depending on the telescope and tracer used, as well as on the spatial scales over which the slope was determined \citep{BenschStutzkiOssenkopf2001}. Thus, we reanalyzed the $\Delta$-variance data in \citet{BenschStutzkiOssenkopf2001}, fitting average power laws over a wider range of scales than considered in \citet{BenschStutzkiOssenkopf2001}, in order to obtain a rough estimate of the slope $\alpha$, less affected by the particular choice of scales for the fit. We find that our reanalysis mostly agrees with the estimates of $\alpha$ in \citet{BenschStutzkiOssenkopf2001}. We see that the overall slope $\alpha$ in the Polaris Flare is of order $-1$, from which we estimate $\sfe\sim0$ based on the model fits of Figure~\ref{fig:slopes} and Equation~(\ref{eq:fitinverse}) in the Polaris Flare, with a formal upper limit of $\sfe<2\%$. This is in good agreement with observations of the Polaris Flare, showing no sign of star formation activity, although the Saxophone region does seem to indicate some trace of gravitational contraction (N.~Schneider 2012, private communication), which, however, has not led to any star formation in the cloud yet.

Entries 6, 8, and 9 are for dust extinction and CO observations of Vela, NGC 1333 in Perseus, and the Coalsack by \citet{SchneiderEtAl2011} and \citet{BenschStutzkiOssenkopf2001}. The values of $\alpha\sim-0.9\pm0.2$ to $-0.8\pm0.3$ were estimated based on average fits to the $\Delta$-variance plots in \citet{SchneiderEtAl2011} and \citet{BenschStutzkiOssenkopf2001}, taking into account a broad range of scales and all the error bars of those analyses. The purpose of our reanalysis is to obtain general trends of the slope $\alpha$, without constraining the fits to relatively small scales as done in \citet{SchneiderEtAl2011} and \citet{BenschStutzkiOssenkopf2001}. This is why we find slightly different slopes here compared to their small-scale fits. These values of $\alpha$ indicate $\sfe\sim0$ with an upper limit of $\sfe<1\%$ in Vela and NGC 1333, and $\sfe<2\%$ in Coalsack. The latter is in agreement with an independent estimate of star formation activity in Coalsack, suggesting relatively low SFE \citep{KainulainenEtAl2009}.

Estimates of $\alpha$ for the following clouds and cloud regions (entries 10, 11, and 13--24) range from $\alpha\sim-0.7\pm0.2$ to $-0.1\pm0.2$, suggesting SFEs in the range $\sfe\sim0$ to a maximum of 6\%, when comparing to the range of SFEs obtained in the simulations for such $\alpha$ (see Figure~\ref{fig:slopes}, Equation~\ref{eq:fitinverse}, and Table~\ref{tab:fits}). Those clouds include Lupus, Rosette, Pipe, Corona Australis, and Perseus, which indeed seem to be consistent with low to moderate star formation activity \citep[compare, e.g., Table~1 in][]{KainulainenEtAl2009}. A notable exception may seem Orion A+B, which contain a large number of YSOs, indicating high star formation activity. In a recent survey of Orion A+B, \citet{MegeathEtAl2012} indeed detected 3479 YSOs. However, Orion A+B are also quite massive with $M_c\sim1.9\times10^5\,\msol$ \citep{WilsonEtAl2005}. Assuming that YSOs produce stars with a mean mass of about $\langle m\rangle\sim0.3\,\msol$ \citep[which is in between the mean masses of the present-day stellar mass function, $\langle m\rangle=0.23\,\msol$, and the initial mass function, $\langle m\rangle=0.37\,\msol$, for disk galaxies; see][]{Chabrier2003}, we can estimate the SFE of Orion A+B with $\sfe\sim3500\times0.3\,\msol/1.9\times10^5\,\msol\sim0.6\%$. A similar estimate based on YSO counts and assuming a median mass for each YSO of $0.5\,\msol$ is provided in \citet{LadaLombardiAlves2010}. From their Table~2, we can compute $\sfe\sim \mathrm{SFR}\times\tsf/M_c$ with an assumed star formation timescale $\tsf=2\,\mathrm{M}\yr$ \citep{EvansEtAl2009,HeidermanEtAl2010,LadaLombardiAlves2010}. Using the values quoted for the low extinction threshold, $\ak\ge0.1\,\mathrm{mag}$, covering almost the whole cloud, we obtain $\sfe\sim2.1\%$ for Orion A and $\sfe\sim0.4\%$ for Orion B, consistent with our own estimate above. Those SFEs obtained from YSO counts, however, are likely lower limits, because of the limited detectability of YSOs on small spatial scales in crowed, dense regions. The YSO completeness achieved with \emph{Spitzer} may only be 50\% averaged over the inner $0.5\,\pc$ of the Trapezium region in Orion, and only 30\% within a radius of $0.2\,\pc$ \citep{EvansEtAl2009,GutermuthEtAl2011}. Thus, SFE estimates based on YSO counts can be quite uncertain. However, they do fall in the range \mbox{$\sfe=0\%$--$4\%$} that we obtain for Orion A+B by using the slope $\alpha$ listed in Table~\ref{tab:obs}. Thus, even if no information on YSOs is available for a given cloud, we can still get a lower and upper limit of the SFE from the slope $\alpha$, which is directly measurable from the column density map of a cloud alone.

Taurus, W3, and Chamaeleon (entries 25--27) all have $\alpha=0.0\pm0.2$, indicating \mbox{$\sfe=0\%$--$10\%$}. For instance, Taurus has a total molecular mass of about $1.5\times10^4\,\msol$ \citep{PinedaEtAl2010}, and more than 300 YSOs were detected \citep[see Table~1 in][]{KainulainenEtAl2009,Elias1978}. Similar to our estimate for Orion A+B above, we find $\sfe\sim300\times0.3\,\msol/1.5\times10^4\,\msol\sim0.6\%$. Another estimate based on YSO counts in \citet{LadaLombardiAlves2010} yields $\sfe\sim1.1\%$ (again using their low extinction threshold value to define $M_c$). Both are consistent with our formal upper limit of $\sfe<10\%$ from $\alpha$, but significantly smaller than the mean, $\sfe\sim5\%$. Taurus is indeed somewhat special, in that it may be more dominated by magnetic fields than other clouds. Recent observations suggest that turbulence in Taurus is trans-Alfv\'enic \citep{HeyerBrunt2012}, leading to lower SFRs than in super-Alfv\'enic clouds (see Paper I). The rather low SFE in Taurus may thus be a result of the relatively strong magnetic fields there. Our estimate of \mbox{$\sfe=0\%$--$10\%$} from the slope $\alpha$ is slightly biased toward larger values of $\sfe$, because stronger magnetic fields also produce slightly larger $\alpha$ (see Figure~\ref{fig:slopes}, bottom panel).

Cloud region 28 in Table~\ref{tab:obs} is a $\Delta$-variance measurement by \citet{SchneiderEtAl2011} of the Cygnus X giant molecular cloud from a high-resolution $^{13}\mathrm{CO}$ integrated intensity map. Cygnus X is one of the most massive and most active star-forming regions in the Galaxy, including both old and young populations of stars \citep{SchneiderEtAl2006}. On well-resolved scales, between $4$ and $20\,\pc$, we find a power-law slope $\alpha\sim0.2\pm0.2$ from Figure~2 in \citet{SchneiderEtAl2011}. Comparing this slope to our simulation models in Figure~\ref{fig:slopes}, we estimate that Cygnus X has an SFE in the range \mbox{$\sfe=0\%$--$17\%$}. The lower limit of the $\sfe$ is formally obtained from the lower limit of $\alpha$, while we actually know that Cygnus X is forming stars, so $\sfe>0$. In particular, the massive DR21 filament in Cygnus X shows signs of infall \citep{SchneiderEtAl2010}, which is likely the reason for Cygnus X exhibiting a slightly positive $\alpha$ on length scales representative of that filament (length $\sim20\,\pc$, width $\sim5\,\pc$) and sub-filaments collapsing toward the main filament \citep{SchneiderEtAl2010,HennemannEtAl2012}. The range \mbox{$\sfe=0\%$--$17\%$} found for Cygnus X is relatively large, but---to the best of our knowledge---the only currently available estimate of the $\sfe$ for the whole Cygnus X giant molecular cloud complex.

Finally, \citet{OssenkopfKlessenHeitsch2001} studied the $\Delta$-variance spectrum of the Serpens core observed by \citet{TestiSargent1998}. \citet{OssenkopfKlessenHeitsch2001} estimated a slope $\alpha=1.6\pm0.4$, which is clearly positive and larger than any of the previous cloud regions studied, indicating $\sfe>8\%$ (last entry in Table~\ref{tab:obs}). Based on the total available gas mass and the mass spectrum of condensations reported in \citet{TestiSargent1998}, we estimate an independent lower limit of the SFE in the Serpens core of $\sfe>5\%$. Recently, \citet{MauryEtAl2011} reported \mbox{$\sfe\sim7\%$--$15\%$} in Serpens South and W40. Another independent estimate by \citet{OlmiTesti2002} suggests an \mbox{$\sfe\sim25\%$--$50\%$} in the region where most protostellar cores and protostars are located. All independent estimates agree with the lower limit, $\sfe>8\%$, that we find here based on $\alpha$. It is indeed plausible that the Serpens core region has a relatively high $\sfe$, because---unlike the large-scale cloud regions studied above---the Serpens core is a small-scale, high-density region, where eventually we expect a relatively large fraction of the gas to be accreted by stars. The Serpens core region is close to the limit where the $\sfe$ approaches the local efficiency $\epsilon$, i.e., the mass fraction of gas of an individual dense core that will actually fall onto the protostar. Theoretical, numerical, and observational estimates suggest that \mbox{$\eps=0.25$--$0.7$} \citep{MatznerMcKee2000,WangEtAl2010,BeutherEtAl2002}. For instance, \citet{WilkingLada1983} find \mbox{$\eps\sim0.3$--$0.5$} and \citet{OlmiTesti2002} find \mbox{$\eps\sim0.25$--$0.5$}. Comparing the SFR and gas column densities in our numerical simulations with Galactic observations by \citet{HeidermanEtAl2010} in Paper I, we found best-fit values \mbox{$\eps=0.3$--$0.7$}, which seem to suggest typical local efficiencies of $\eps\sim0.5$, in good agreement with the aforementioned independent estimates. Thus, about half of the gas in a dense core is accreted by the protostar, while the other half remains in the envelope or is driven out of the core by winds, jets, and outflows.

\section{Uncertainties and Limitations} \label{sec:uncertainties}

\subsection{Relation between $\alpha$ and $\sfe$}
In Section~\ref{sec:estimatingsfe} we calibrated a simple, empirical relation given by Equation~(\ref{eq:fitinverse}) between the spectral slope $\alpha$ of the column density and the $\sfe$ in the numerical simulations of Section~\ref{sec:sims}. This relation can be applied to observational maps in which $\alpha$ can be measured with Fourier techniques such as the $\Delta$-variance method, to obtain an estimate of the $\sfe$ in the observed region. Equation~(\ref{eq:fitinverse}) only includes an uncertainty in $\alpha\pm0.2$ as suggested by our calibration with numerical simulations. The practical uncertainties in $\alpha$ when measured in an observational map, however, can be higher than that. Such uncertainties include finite telescope resolution, signal-to-noise, and sensitivity limits for a given observational tracer of column density. For instance, dust emission, extinction, or molecular line emission all suffer from sensitivity limits in both the high and low column-density regimes (e.g., molecular lines become optically thick at high column density or detectors are simply not sensitive enough to trace very low column density). Such statistical and systematic uncertainties are not included in Equation~(\ref{eq:fitinverse}) and must be added accordingly.

\subsection{Global versus Local Efficiencies}
A primary goal of this paper is to understand the transition from the global $\sfe$ on large scales (e.g., a whole molecular cloud) to the local core formation efficiency $\eps$. The latter denotes the fraction of gas in a dense core that eventually goes into the star. This is not 100\%, because some fraction, $1-\eps$, of the gas in the core does not fall onto the protostar, but is driven out of the core by outflows, winds, and/or jets during the formation of the protostar in the center of the accretion disk. Without specifying what mechanisms (e.g., radiative and/or magnetic) cause the gas flow out of the core or how it actually happens, we simply note that we expect a local efficiency \mbox{$\eps\approx0.3$--$0.7$} \citep[][Paper I, and references therein]{MatznerMcKee2000}. However, even the best high-resolution observations of individual cores are just beginning to investigate this local regime of star formation and constrain $\eps$. In surveys of structure and star formation activity of molecular clouds on intermediate and large scales, such cores appear as sites of high column density, but their internal structure is often not resolved. Thus, the actual amount of gas contributing to star formation is uncertain and so is the actual total $\sfe$. We note that we did not take into account the local efficiency $\eps$ in Equation~(\ref{eq:fitinverse}), because the simulations did not include any form of feedback such as winds, jets, or outflows from the protostellar objects. Such feedback effects are not accounted for and hence add to the uncertainty in the true $\sfe$ when Equation~(\ref{eq:fitinverse}) is applied to observational data.

Given the local efficiencies \mbox{$\eps\approx0.3$--$0.7$}, it would seem reasonable to adjust the inferred SFE via Equation~(\ref{eq:fitinverse}) down by the factor $\eps$ to estimate the actual total SFE of a cloud. In the application of Equation~(\ref{eq:fitinverse}) to the observations presented in Section~\ref{sec:obs}, however, we did not make such an adjustment, because we think that the same adjustment would be necessary in the case of YSO counts to infer the SFE. This is because the effect of the local efficiency $\eps$ is typically also not accounted for when estimating the typical mass of a YSO, since those observations are limited in resolution and likely include a significant fraction of gas/dust in the core that may not all be accreted by the star. Thus, the local efficiency $\eps$ due to feedback introduces uncertainties as discussed above, which, however, similarly apply to SFE estimates obtained with YSO counts.

\subsection{Definition of the SFE in Closed and Open Systems}
We furthermore note that the simulations studied here are highly idealized and can suffer from resolution issues. However, the effects of limited resolution are relatively small for our measurement of $\alpha$ and we do understand them quite well (see Appendix~\ref{app:spectres}). A more severe issue is the choice of boundary conditions in the simulations. They are periodic in all directions. We discuss the limitations of this and related, different approaches to treat boundaries in Paper I. For the present study of the SFE, the main limitation lies in the fact that our computational boxes are closed, so there is no gas inflow or outflow. Real molecular clouds are not isolated, closed objects, but exchange mass with the medium in which they are embedded. This poses a general problem and uncertainty in defining the $\sfe$. Imagine, for instance, a cloud accreting more gas from the outside than is turned into stars inside in the same amount of time. In that case, the $\sfe$ would go down, even though stars form continuously, simply because the total gas content in the region of interest increases due to accretion from the outside. An even more remarkable thought experiment is one in which star formation has stopped at a given $\sfe$ in a given volume, and suppose the stars formed in that region now blow away all the remaining gas in their surrounding and drive it out of that volume (e.g., by driving winds). According to the standard definition of the $\sfe$ in Equation~(\ref{eq:sfe}), $\mstar$ would remain the same in this process, but $\mgas$ would go down, leading to an increasing $\sfe$, even though no more stars form. This is an example emphasizing that one must be quite careful with interpreting SFEs in open systems. Such definitions of the $\sfe$ in open systems are discussed in \citet{VazquezSemadeniEtAl2010} and \citet{FeldmannGnedin2011}.

\section{Summary and Conclusions} \label{sec:conclusions}

We studied the star formation efficiency (SFE) in simulations of supersonic, self-gravitating, magnetized turbulence. In Paper I, we found that the simulations yield star formation rates (SFRs) varying by orders of magnitude, depending on the virial parameter, on the forcing of the turbulence, and on the sonic and Alfv\'en Mach numbers, consistent with observed SFR column densities. Here, we focused on the SFE and find that the density and column density distributions change significantly when star formation occurs and when the SFE increases. We find that the column density scaling is particularly sensitive to the SFE, enabling us to estimate the SFE from column density observations, without requiring a priori information on star formation activity or YSO counts. More detailed results are listed in the following.

\begin{enumerate}

\item{We find that star formation affects the column density structure of clouds (see Figures~\ref{fig:imagessol} and~\ref{fig:imagescomp}). The effect of star formation on the global structure is stronger in low Mach number clouds and for clouds with solenoidal forcing than for clouds with high Mach number and/or compressive forcing, because strong turbulence alone produces high-density filaments, without requiring gravitational contraction.}

\item{The volumetric and column density PDFs become wider with increasing Mach number and more compressive forcing (see Figures~\ref{fig:pdfs} and~\ref{fig:coldenspdfs}). In contrast, increasing the magnetic field strength yields smaller standard deviations of the PDF (see Figure~\ref{fig:magpdfs}), as expected from Equation~(\ref{eq:sigs}).}

\item{The PDFs of our simulations with extremely different forcing, sonic Mach number, and magnetic field strengths, all develop power-law tails of flattening slope with increasing SFE. We find that the high-density tails of the PDFs are consistent with equivalent radial density profiles, $\rho\propto r^{-\kappa}$ with \mbox{$\kappa=1.5$--$2.5$} depending on $\sfe$, consistent with column density observations \citep{MotteAndreNeri1998,ShirleyEtAl2000,KainulainenEtAl2009,ArzoumanianEtAl2011,SchneiderEtAl2012}.}

\item{The velocity spectra from our simulations with different forcing, sonic Mach number, and magnetic field are all consistent with the scaling inferred in molecular cloud observations, \mbox{$v\propto\ell^{1/2}$} \citep{Larson1981,SolomonEtAl1987,FalgaronePugetPerault1992,OssenkopfMacLow2002,HeyerBrunt2004,RomanDuvalEtAl2011}}, suggesting highly compressible, supersonic turbulence of Burgers type on large scales. We find evidence for a transition to subsonic, Kolmogorov turbulence with \mbox{$v\propto\ell^{1/3}$} on scales smaller than the sonic scale (see Figure~\ref{fig:velspectra}). The velocity spectra are largely insensitive to star formation.

\item{Unlike the velocity spectra, we find a strong change of the density and column density spectra with the onset of star formation. The slope $\alpha$ of $\pthreed(\rho,k)\propto k^\alpha$ or equivalently of the $\Delta$-variance spectrum of column density, $\dvar(\Sigma,\ell)\propto\ell^{-\alpha}$, which are both related to the column density spectrum $\ptwod(\Sigma,k)$ via Equations~(\ref{eq:ptwodtothreedrelation}) and~(\ref{eq:dvarspectrelation}), switches sign from $\alpha<0$ for $\sfe\sim0$ to $\alpha>0$ when star formation proceeds, $\sfe\gtrsim0$ (see Figures~\ref{fig:coldensspectra} and~\ref{fig:Bspectra}).}

\item{The change in slope $\alpha$ occurs in all our models, despite the wide range of physical cloud conditions probed in the simulations (\mbox{$\mach\sim3$--$50$} with solenoidal, mixed, and compressive forcing of the turbulence, and magnetic fields ranging from the super-Alfv\'enic to the trans-Alfv\'enic regime; see Table~\ref{tab:sims}). We confirmed this result for the slope $\alpha$ with an independent measurement based on the $\Delta$-variance technique, suitable for molecular cloud observations (see Figures~\ref{fig:deltavar} and~\ref{fig:Bdeltavar}).}

\item{Studying the detailed dependence of the slope $\alpha$ on $\sfe$ in all our numerical models, we find that the $\sfe$ can be estimated by measuring $\alpha$ (see Figure~\ref{fig:slopes}). We provide a fitted relation, $\sfe(\alpha)$ in Equation~(\ref{eq:fitinverse}) for each numerical model set with best-fit parameters listed in Table~\ref{tab:fits}.}

\item{We compared observational measurements of $\alpha$ in an $\mathrm{H}\,\textsc{i}$ cloud, in several giant molecular clouds and sub-regions, and in a dense star-forming core, with values ranging from $\alpha=-1.6$ to $+1.6$, all listed in Table~\ref{tab:obs}. We infer the $\sfe$ for each cloud or region studied and show that negative $\alpha$ typically indicates $\sfe\sim0$, consistent with observations. Clouds with $\alpha\sim0$ and larger do form stars and exhibit a range of SFEs, which is monotonically increasing with $\alpha$. Estimates of $\sfe$ from $\alpha$ are consistent with independent measurements based on YSO counts, where available. However, the $\sfe(\alpha)$ relations found here provide an independent estimate of $\sfe$ based on a column density map alone, without requiring a priori knowledge of star-formation activity or YSO counts for a given cloud.}

\end{enumerate}

The overall agreement between our simulations and observations of Galactic clouds concerning the SFE studied here and the SFR in Paper I is promising. We conclude that supersonic, magnetized turbulence is a key process, likely controlling the SFRs and SFEs of molecular clouds in the Milky Way and potentially in other galaxies.

\acknowledgements
We thank Nicola Schneider, Timea Csengeri, J\"org Fischera, Jouni Kainulainen, and Volker Ossenkopf for useful discussions on column density PDFs and $\Delta$-variance measurements in Galactic clouds. The work on PDF tails was partly inspired by discussions with Philipp Girichidis and Lukas Konstandin. We thank the anonymous referee for suggesting a critical discussion of the uncertainties in estimating the SFE in observations.
C.~F.~acknowledges the Australian Research Council for a Discovery Projects Fellowship (grant No.~DP110102191). R.~S.~K.~acknowledges subsidies from the Baden-W\"urttemberg-Stiftung via contract research grant P-LS-SPII/18.
This work was supported by the Deutsche Forschungsgemeinschaft, priority program 1573 ("Physics of the Interstellar Medium") and the collaborative research project SFB 881 ("The Milky Way system").
The simulations presented in this work were run on supercomputers at the Leibniz Rechenzentrum (project pr32lo) and at the Forschungszentrum J\"ulich (project hhd20).
The software used here was in part developed by the DOE-supported ASC/Alliance Center for Astrophysical Thermonuclear Flashes at the University of Chicago.

\appendix

\section{Dependence of the Density PDFs on Resolution} \label{app:pdfres}

\begin{figure*}[t]
\centerline{
\includegraphics[width=0.95\linewidth]{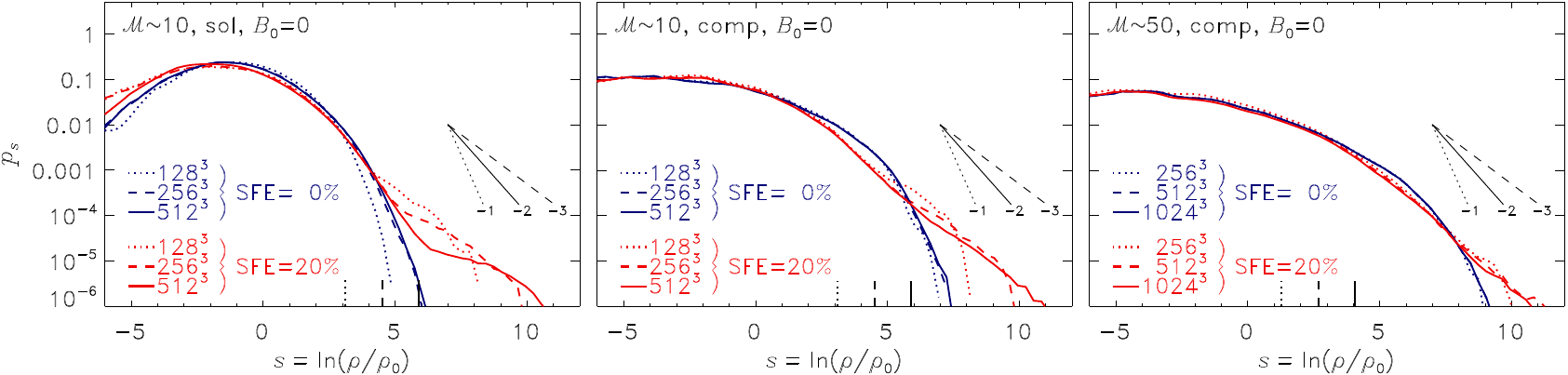}
}
\caption{Same as Figure~\ref{fig:pdfs}, but for a study of the influence of the numerical resolution on the density PDF for $\sfe=0$ and $\sfe=20\%$ in three representative models. \emph{Left}: solenoidal forcing with $\mach\sim10$ at resolutions of $128^3$, $256^3$, and $512^3$ grid cells. \emph{Middle}: same as left panel, but for compressive forcing with $\mach\sim10$. \emph{Right}: compressive forcing with $\mach\sim50$ at resolutions of $256^3$, $512^3$, and $1024^3$. Model parameters are listed in Table~\ref{tab:sims}.}
\label{fig:pdfres}
\end{figure*}

Figure~\ref{fig:pdfres} shows a resolution study of the density PDF $p(s)$ for our two Mach 10 models with solenoidal and compressive forcing (left and middle panel), and for our Mach 50 model with compressive forcing (right panel) at $\sfe=0$ and $\sfe=20\%$. We compare resolutions $N_\mathrm{res}=128$, $256$, and $512$ for the Mach 10 models. We see no significant resolution dependence for $N_\mathrm{res}\ge256$ in the log-normal part of the PDF, but a clear resolution dependence in the power-law tail, stretching the PDF to higher $s$ when the resolution is increased. In the solenoidal forcing case (left panel), we find a break in the power law at $s_\mathrm{break}\sim4.0$, 5.2, and 6.5 for $N_\mathrm{res}=128$, 256, and 512, respectively. These breaks seem to correlate with the density threshold for sink particle creation, Equation~(\ref{eq:rhosink}), $s_\mathrm{sink}=\ln(\rho_\mathrm{sink}/\meanrho)\sim3.1$, 4.5, and 5.9, shown as vertical short lines on top of the abscissa ($N_\mathrm{res}=128$:~dotted; 256:~dashed, and 512:~solid), but are slightly higher than that. The PDF for densities $s<s_\mathrm{break}$ seems converged, while the flatter part for which $s>s_\mathrm{break}$ shifts to higher densities when the resolution is increased. The compressive forcing model with Mach 10 (middle panel) does not show such a break, but a cutoff at large $s$, which shifts to higher $s$ with increasing resolution, similar to the cutoff seen in the solenoidal forcing model at large $s$. The PDF of the Mach 50, compressive forcing case (right panel) does not show a significant resolution dependence for the range of densities and $p_s$ analyzed here ($N_\mathrm{res}=256$:~dotted; 512:~dashed; and 1024:~solid), even though the sink particle creation densities are quite low (see the vertical lines on top of the abscissa).

\section{Dependence of the Fourier Spectra on Resolution} \label{app:spectres}

\begin{figure*}[t]
\centerline{
\includegraphics[width=0.95\linewidth]{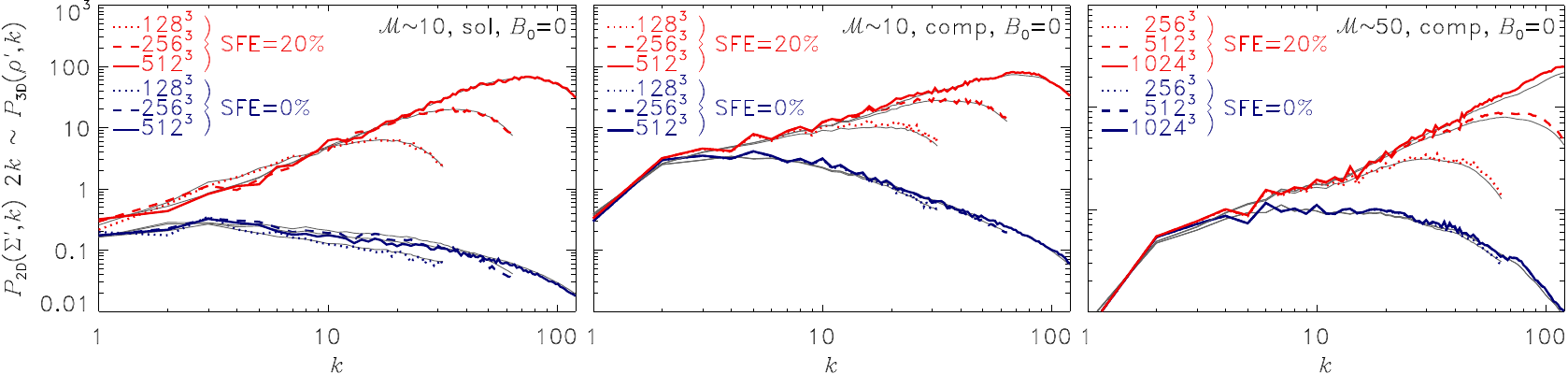}
}
\caption{Same as Figure~\ref{fig:pdfres}, but for a resolution study of the density and column density Fourier spectra as in Figure~\ref{fig:coldensspectra}. The significant change in slope $\alpha$ with $\sfe$ is clearly seen, even for our lowest numerical resolution.}
\label{fig:coldensspectra_res}
\end{figure*}

Figure~\ref{fig:coldensspectra_res} shows the same resolution study as in Figure~\ref{fig:pdfres}, but for the density and column density power spectra as in Figure~\ref{fig:coldensspectra}. This resolution study demonstrates that density and column density spectra are converged for $k \lesssim N_\mathrm{res} / 15$. In contrast, velocity spectra are only converged for $k \lesssim N_\mathrm{res} / 30$ \citep{FederrathDuvalKlessenSchmidtMacLow2010,FederrathSurSchleicherBanerjeeKlessen2011}. This is because resolving the kinetic energy content of a vortex (which is a vector quantity) requires at least 30 grid cells. Here we find that about 15 grid cells are sufficient to resolve turbulent and collapsing density structures. This in turn suggests that resolving the Jeans length with 15 grid cells would be sufficient to capture the density structures in collapse simulations, but is insufficient to resolve kinetic vorticity and potential small-scale dynamo amplification of magnetic fields (both of which are vector quantities), which still requires at least 30 grid cells per Jeans length \citep{SurEtAl2010,FederrathSurSchleicherBanerjeeKlessen2011}. The slopes $\alpha$ that we estimate in \mbox{Figures~\ref{fig:coldensspectra}--\ref{fig:Bdeltavar}} are obtained in a converged range of scales, and hence these slopes are not affected by numerical resolution.

\section{Dependence of the Density PDFs and Fourier Spectra on the Random Seed} \label{app:seeds}

\begin{figure*}[t]
\centerline{
\includegraphics[width=0.75\linewidth]{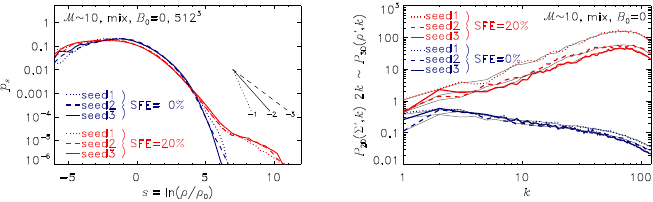}
}
\caption{\textit{Left}: same as Figure~\ref{fig:pdfres}, but for a study of the influence of different random seeds of the turbulence (seed1, seed2, seed3) on the PDF for $\sfe=0$ and $\sfe=20\%$ in three mixed-forcing runs with $\mach\sim10$ and $512^3$ resolution (see Table~\ref{tab:sims}). \textit{Right}: same as left panel, but for the density and column density spectra as in Figure~\ref{fig:coldensspectra_res}.}
\label{fig:pdfspectseeds}
\end{figure*}

Figure~\ref{fig:pdfspectseeds} (left) indicates statistical fluctuations mostly in the low- and high-density tails of the PDFs when different random seeds of the turbulence (seed1, seed2, seed3) are considered. This gives us an impression of the statistical uncertainties of the PDFs. The right panel shows the density and column density spectra of the same models, also indicating statistical fluctuations, which manifest in different amplitudes of the spectra, but the power-law slopes do not vary significantly with random seed.


\end{document}